\documentclass[11pt, a4paper]{article}
\pdfoutput=1

\usepackage{jcappub}

\usepackage{comment}
\usepackage{amsmath}
\usepackage{amsfonts}
\usepackage{amssymb}
\usepackage{graphicx}
\usepackage{mathrsfs}
\usepackage{latexsym}
\usepackage{color}
\usepackage{slashed, cancel}
\usepackage{hyperref}
\usepackage{cleveref}
\usepackage{float}
\usepackage{tikz}
\usepackage{bbold}
\usepackage{soul} 
\usepackage{mathtools}
\usepackage{support-caption} 
\usepackage{subcaption}
\usepackage{enumitem}
\usepackage{journals}
\allowdisplaybreaks

\usepackage[utf8]{inputenc} 

\hypersetup{urlcolor=blue, colorlinks=true} 

\usepackage{fancyhdr}
\renewcommand{\arraystretch}{1}
\numberwithin{equation}{section}

\definecolor{rossos}{rgb}{0.8,0.2,0.3}
\definecolor{bluscuro}{rgb}{0.15, 0.2, .85}
\definecolor{bluchiaro}{cmyk}{1,.3,0.,0.1}
\hypersetup{colorlinks, citecolor=bluscuro, linkcolor=bluscuro, urlcolor=bluscuro}

\makeatother   
\newcommand{\GeV}{{\rm \,GeV}}
\newcommand{\TeV}{{\rm \,TeV}}
\newcommand{\MeV}{{\rm \,MeV}}

\newcommand{\cm}{{\rm \,cm}}
\newcommand{\km}{{\rm \,km}}
\newcommand{\s}{{\rm \,s}}

\newcommand{\alphaem}{\alpha_{\rm em}}

\newcommand{\erf}{{\rm \,Erf}}

 \def\be   {\begin{equation}}   \def\ee   {\end{equation}}
 \def\ba   {\begin{array}}      \def\ea   {\end{array}}
 \def\bea  {\begin{eqnarray}}   \def\eea  {\end{eqnarray}}
 \def\bean {\begin{eqnarray*}}  \def\eean {\end{eqnarray*}}

\begin{document}

\today

\title{Capture of Leptophilic Dark Matter in Neutron Stars}

\author[a]{Nicole F.\ Bell,}
\author[b]{Giorgio Busoni}
\author[a]{and Sandra Robles}
\affiliation[a]{ARC Centre of Excellence for Particle Physics at the Terascale \\
School of Physics, The University of Melbourne, Victoria 3010, Australia}
\affiliation[b]{Max-Planck-Institut fur Kernphysik, Saupfercheckweg 1, 69117 Heidelberg, Germany.}

\emailAdd{\tt n.bell@unimelb.edu.au}
\emailAdd{\tt giorgio.busoni@mpi-hd.mpg.de}
\emailAdd{\tt sandra.robles@unimelb.edu.au}

\abstract{
Dark matter particles will be captured in neutron stars if they undergo scattering interactions with nucleons or leptons. These collisions transfer the dark matter kinetic energy to the star, resulting in appreciable heating that is potentially observable by forthcoming infrared telescopes. While previous work considered scattering only on nucleons, neutron stars contain small abundances of other particle species, including electrons and muons. We perform a detailed analysis of the neutron star kinetic heating constraints on leptophilic dark matter. We also estimate the size of loop induced couplings to quarks, arising from the exchange of photons and Z bosons. Despite having relatively small lepton abundances, we find that an observation of an old, cold, neutron star would provide very strong limits on dark matter interactions with leptons, with the greatest reach arising from scattering off muons. The projected sensitivity is orders of magnitude more powerful than current dark matter-electron scattering bounds from terrestrial direct detection experiments. 
}

\maketitle

\section{Introduction}

The capture of dark matter (DM) in stars is a well-established method to probe the particle properties of dark matter.  While the direct detection (DD) of dark matter particles in terrestrial experiments remains elusive, these “cosmic laboratories” provide an important means of exploring dark matter interactions with regular matter and narrowing down the plethora of possible dark matter candidates. In doing so, they can provide much needed guidance for the targeting of DM direct detection experiments.

If ambient dark matter particles scatter with ordinary matter in stars, they can lose kinetic energy and become gravitationally captured by the star~\citep{Gould:1987ju,Gould:1987ir,Jungman:1995df,Kumar:2012uh,Kappl:2011kz,Busoni:2013kaa,Bramante:2017xlb}. Over time, this leads to an accumulation of dark matter in the core of the star, which can have observable consequences. For example, the accumulation and subsequent annihilation of dark matter particles in the Sun can produce highly energetic fluxes of neutrinos that could potentially be seen by neutrino telescopes, providing an important means of dark matter indirect detection \cite{Tanaka:2011uf,Choi:2015ara,Adrian-Martinez:2016ujo,Adrian-Martinez:2016gti,Aartsen:2016zhm}. Alternatively, the annihilation of captured dark matter to long-lived dark mediators can lead to spectacular signals~\cite{Batell:2009zp,Schuster:2009au,Bell:2011sn,Feng:2016ijc,Leane:2017vag}.

If we instead consider a neutron star (NS), we can constrain the type and strength of dark matter interactions via the requirements that the resultant neutron star heating is not too large~\citep{Baryakhtar:2017dbj,Raj:2017wrv,Gonzalez,Bell:2018pkk,Camargo:2019wou}, that the star does not accumulate so much dark matter that collapse to a black hole is triggered~\citep{Kouvaris:2010jy,Kouvaris:2011fi,McDermott:2011jp,Guver:2012ba,Bell:2013xk, Bramante:2013nma,Bertoni:2013bsa,Garani:2018kkd}, and that the neutron star structure is not perturbed to the extent that the gravitational wave signatures from binary neutron star mergers are inconsistent with observations~\citep{Ellis:2017jgp,Ellis:2018bkr,Nelson:2018xtr}.

Neutron stars provide particularly powerful DM constraints, because their high density leads to efficient capture. Indeed, for DM-nucleon cross sections above a threshold value of order $10^{-45}$ cm${^2}$, the capture probably saturates at the geometric limit, $\sigma \sim \pi R_*^2 m_n/M_*$, such that all dark matter incident on the star is captured. This conclusion holds irrespective of whether DM scattering interactions are spin independent (SI) or spin dependent (SD). Given that conventional direct detection experiments currently have sensitivity to DM-nucleon cross sections below $10^{-45}$ cm${^2}$ only in a limited DM mass range, and only for SI  interactions, NS techniques are clearly very useful. Moreover, velocity or momentum dependent interactions are completely inaccessible to terrestrial direct detection experiments, being severely suppressed in the non-relativistic regime applicable for scattering on Earth. In comparison, DM particles are accelerated to quasi-relativistic velocities upon NS infall, effectively erasing such kinematic suppression.

When dark matter is gravitationally captured by a NS, the kinetic energy transferred in the collisions heat up the star. The captured dark matter will then undergo a series of further collisions, eventually transferring almost all of its initial kinetic energy, to reach a state of thermal equilibrium with the star. As shown in Refs.~\citep{Baryakhtar:2017dbj,Raj:2017wrv} this can heat neutron stars up to 1700K.\footnote{In addition to the heating from capture, there may also be heating from the annihilation of the captured dark matter, leading to an additional temperature increase of order 700K~\citep{Baryakhtar:2017dbj}. This annihilation heating has not been included in our calculations as it is more model dependent, e.g., asymmetric dark matter would not annihilate (see however Refs.~\citep{Baldes:2017gzu,Baldes:2017gzw}) or the dark matter may not fully thermalize.}
Because old isolated neutron stars can cool to temperatures below 1000K, such heating (or its absence) can be used to place limits on the strength of dark matter interactions.\footnote{Other potential sources of heating include rotochemical heating, recently discussed in \citep{Hamaguchi:2019oev}.}  
Importantly, this kinetic heating may be within reach of forthcoming infrared telescopes~\citep{Baryakhtar:2017dbj,Raj:2017wrv}. Provided NSs are nearby, faint and sufficiently isolated, they are likely to be discovered by existing radio telescopes such as the Five-hundred-meter Aperture Spherical radio Telescope (FAST)~\citep{Nan2011}, or the future Square Kilometer Array (SKA)~\citep{Konar:2016lgc}. Their thermal emission can then be measured by infrared telescopes such as the James Webb Space Telescope (JWST), the Thirty Meter Telescope (TMT), or the European Extremely Large Telescope (E-ELT) \cite{Baryakhtar:2017dbj}.

The projected NS kinetic heating limits on the DM-nucleon interaction strength were recently calculated in~\citep{Baryakhtar:2017dbj,Raj:2017wrv,Bell:2018pkk}, where they were found to be comparable to, or in many cases much stronger than, existing or projected limits from direct detection experiments.  Specifically, xenon-based DD experiments can provide superior limits only for unsuppressed SI scattering (i.e., for scalar or vector interactions) and only in the mass range 10~GeV~$< m_\chi <$~1~TeV. Otherwise, the projected NS kinetic heating sensitivity can always compete with or exceed conventional DD techniques, often by several orders of magnitude. Moreover, the quasi-relativitic speed of dark matter particles upon NS infall permits enhanced sensitivity to inelastic dark matter interactions~\citep{Bell:2018pkk}, for which low energy scattering on Earth is suppressed or forbidden.

Previous work on NS kinetic heating considered only dark matter-nucleon scattering. However, neutron star cores contain multiple particle species, including neutrons, protons, electrons and muons, with relative fractions that vary with the NS radius and depend upon the NS equation of state. The purpose of the present paper is determine the sensitivity of NS kinetic heating to {\it DM-lepton scattering cross sections} arising from the interaction of {\it leptophilic DM}~\citep{Kopp:2009et,DEramo:2017zqw, Fox:2011fx,Bell:2014tta,Garani:2017jcj} with the electron and muon components of the star. 
We shall see that, while current direct detection bounds on DM-electron scattering are modest and have greatest sensitivity in the 1-1000 MeV mass range, NS techniques can provide very powerful limits which span a wide range of dark matter masses. 
It is worth mentioning that limits on leptophilic DM models from different astrophysical phenomena, the internal heat flux of Earth and colliders, though less stringent, have been previously derived in the literature (see e.g. refs.~\cite{Guha:2018mli,Chauhan:2016joa,Bertuzzo:2017lwt}). 

One may ask if it is self consistent to consider DM interactions that are purely leptophilic. Indeed, the presence of DM-nucleon couplings are expected at loop level, even if they are absent at lowest order.  As such, it is important to consider whether the strongest bounds arise from tree level DM-lepton scattering or, in fact, from loop induced DM-nucleon scattering.  While such issues have been addressed in the past~\citep{Kopp:2009et,DEramo:2017zqw}, they have always been analysed in the context of non-relativistic scattering, and must therefore be re-examined in the NS context. For many coupling types, we shall see that the tree-level electron and muon interactions will indeed dominate the observable signals.

The organisation of this paper is as follows. We detail the neutron star composition parameters required for our study and outline the DM capture process and its implications for the NS temperature in Section~\ref{sec:NSleptons}. 
We give the relevant expressions to calculate the elastic scattering cross section off leptons for the relativistic regime in Section~\ref{sec:leptons}. 
In Section ~\ref{sec:loops} we estimate the size of loop-induced nucleon couplings, to enable a self-consistent comparison of lepton and nucleon scattering limits. 
In Section~\ref{sec:results} we determine NS kinetic heating constraints on DM-lepton interactions.
Our conclusions can be found in Section \ref{sec:conclusions}.

\section{Leptophilic DM capture in neutron stars}
\label{sec:NSleptons}

We now outline the neutron star properties relevant for dark matter scattering, and hence determine the capture rate for scattering of DM from the different species in a neutron star. 

\subsection{Leptons in neutron stars}

Below a thin atmosphere, neutron stars are usually divided into two concentric regions, a thin crust and a massive core \cite{Haensel:2007yy}. The crust constitutes $\sim 1\%$ of the NS mass and is about 1 km thick. The crust is further divided into the outer and inner crust. The outer crust or envelope, where the mass density $\rho$ is below the neutron drip density $\rho_{ND}\sim(4-6) \times 10^{11} {\rm g \, cm^{-3}}$, contains mainly atomic nuclei and strongly degenerate electrons. In the inner crust, $\rho_{ND} \leq \rho \leq \rho_{cc}$, where $\rho_{cc}$ is the density at the crust-core boundary and is order half the nuclear saturation density, $\rho_0=2.8 \times 10^{14} {\rm g \, cm^{-3}}$, matter  consists of ultrarelativistic highly degenerate electrons, very neutron rich nuclei and free degenerate neutrons (dripped off the nuclei). 
  
The NS core can  be divided into outer and inner core.   
At $\rho_{cc}$, due to inverse beta decay, the nuclear matter dissolves into a uniform liquid composed primarily of strongly degenerate neutrons plus an equal fraction of protons and leptons (electrons and muons)  with abundances in the 5-10\% range \cite{Baym:2006rq,Sedrakian:2006mq}.  Muons appear in the NS core when the electron Fermi energy exceeds the muon mass. This phase, the outer core,  extends to densities of  $\sim 2 \rho_0$ \cite{Yakovlev:2015vma}. The inner core, $\rho\gtrsim 2\rho_0$, extends to the stellar centre, where $\rho \sim 10\rho_0$, depending on the NS mass. The inner core composition remains uncertain and may be the same as in the outer core or essentially different. It may contain hyperons, pion or kaon condensates, or free quarks, or a mixture of these. The proton fraction in the inner core is expected to be larger than in the outer core ($\sim 11-13\%$)~\cite{Yakovlev:2007vs}. 

\begin{figure}[t] 
\centering
\includegraphics[width=0.485\textwidth]{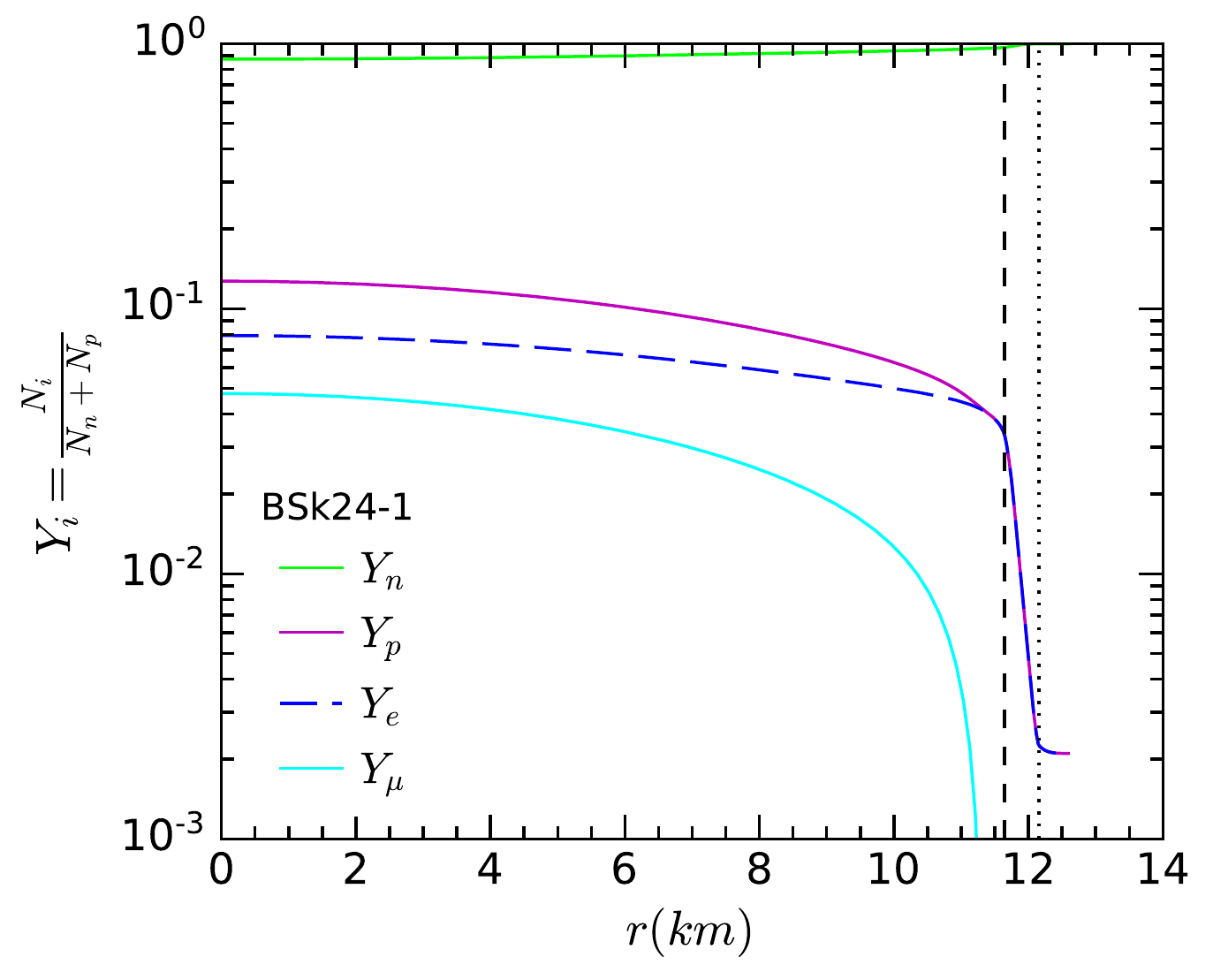}
\includegraphics[width=0.475\textwidth]{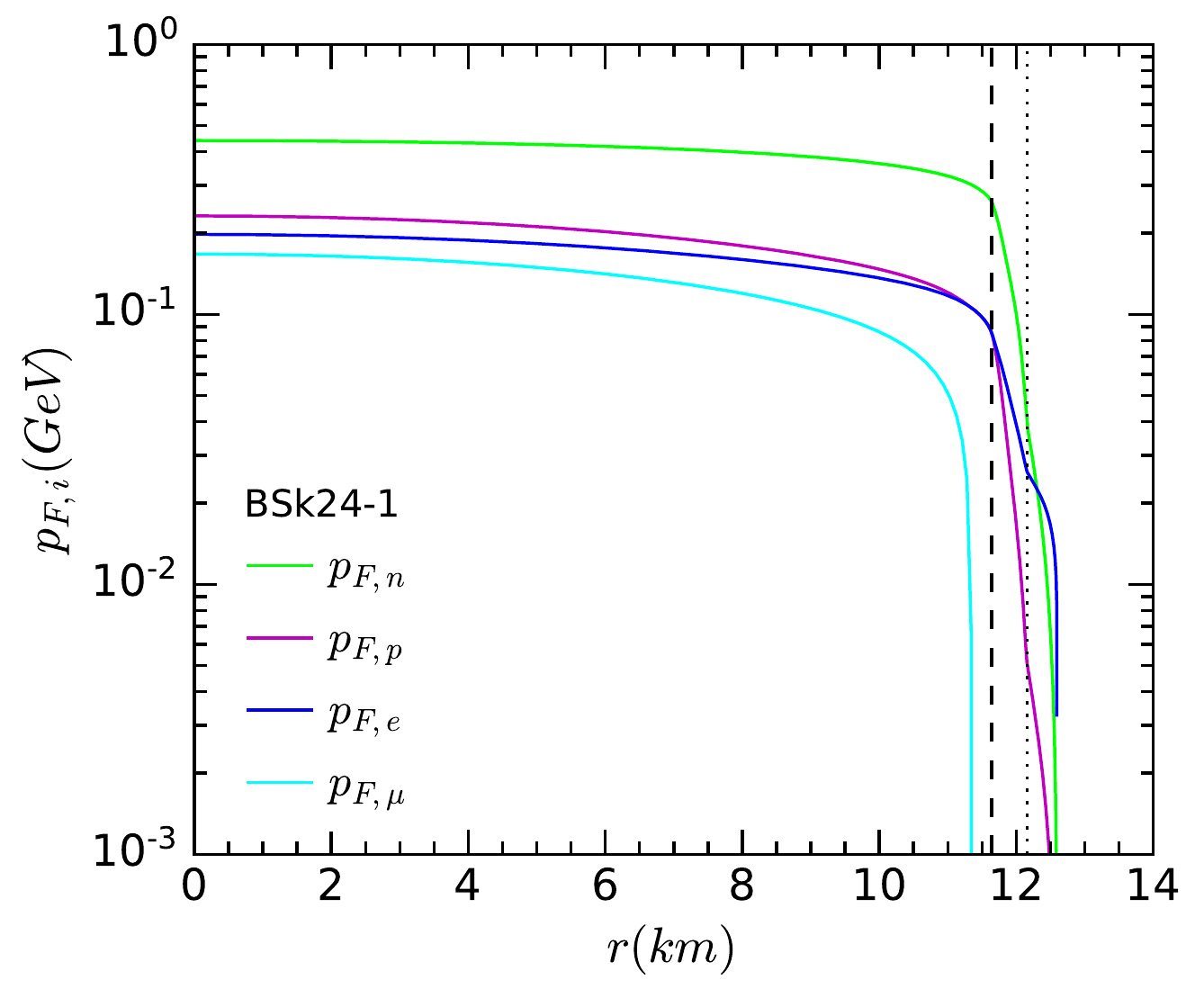}
\caption{Left: $Y_i$ abundances as a function of the NS radius for the different species in a NS with EoS BSk24-1, computed with respect to the baryon number ($N_n+N_p$). Right: Fermi momentum for the same EoS and same species vs radius. The black dashed line denotes the crust-core interface and the dotted line indicates the inner-outer crust boundary.}
\label{fig:YipF}
\end{figure}

As previously outlined, the exact microscopic composition of NSs depends on the number density radial profile. To precisely determine the internal structure of a NS and hence the neutron, proton, electron and muon abundances, $Y_i$, and other microscopic properties such as the Fermi momentum of each species, the equation of state (EoS) of dense matter is a key ingredient. There is a wide range of allowed EoSs, see e.g.~\citep{Annala:2019eax}.
In the left hand side of Figure~\ref{fig:YipF}, we show the radial profile of the particle abundances, $Y_i=N_i/N_b$, where $N_b=N_n+N_p$ is the total number of baryons, $N_n$ and $N_p$ are the total number of neutrons and protons, respectively, in a NS with unified EoS\footnote{A unified equation of state is valid in all the regions of a NS.} BSk24~\cite{Goriely:2013,Pearson:2018tkr} and low mass configuration (BSk24-1)\footnote{Charge neutrality requires $Y_p=Y_e+Y_\mu$. Note also that $Y_n=1 -Y_p$.}. 
We have evaluated all the EoS dependent  quantities that appear in this paper with the parametrizations given in Appendix C of  ref.~\cite{Pearson:2018tkr} and the publicly available {\tt FORTRAN} subroutines implemented by the same  authors\footnote{\url{http://www.ioffe.ru/astro/NSG/BSk/}}. 
Note that the abundances and the Fermi momentum of each species do not vary much along the core; their radial profile is almost flat, except when they approach the crust-core interface (black dashed line in Figure~\ref{fig:YipF}). 
Since our aim is to demonstrate that NS kinetic heating can constrain leptophilic DM models, and the lepton density in the crust is much lower than in the core, we will restrict our analysis to the NS core. 

We will hereafter adopt the NS model BSk24-1 as our benchmark NS in order to derive conservative limits on leptophilic DM models, and use the relevant NS  microscopic properties  of this model averaged over the NS core volume. \footnote{In subsection \ref{sec:eos}, we evaluate the impact of choosing a different NS model.}

\subsection{Capture rate and kinetic heating}
While stars orbit around the centre of the Galaxy, they move through large fluxes of DM particles. 
When DM interacts with Standard Model (SM) particles inside stars, it can lose energy in the scattering and, provided the energy loss is large enough, becomes gravitationally bound to the star. 
The high density of NSs enhances this capture process. In addition, the fact that  the gravitational pull of these stars accelerates DM particles to velocities comparable to the speed of light compel us to take into account relativistic effects.

Considering that DM scatters off a single SM particle within the NS core and the scattering cross section, $\sigma$, is sufficiently large that all DM particles are captured as they transit a NS, $\sigma \gtrsim \sigma_{th}$,  neglecting thermal effects, the capture rate tends to the geometric limit~\cite{Bell:2018pkk},
\begin{equation}
C_{\star} =  \frac{\pi R_{core}^2 (1-B)}{v_\star B} \frac{\rho_\chi}{m_\chi} \erf\left(\sqrt{\frac{3}{2}}\frac{v_\star}{v_d}\right),
\label{eq:Cstar}
\end{equation}
where $\rho_\chi$ is the local DM density, and $B$ is defined as
\begin{equation}
B = 1-\frac{2GM_{core}}{c^2 R_{core}}.    
\end{equation}
We have assumed a Maxwell-Boltzmann distribution for the DM speed, with $v_d$ the velocity dispersion and $v_\star$ the NS speed, which we assume to be comparable to the speed of the Sun.

The threshold cross section, that is the value for which the cross section becomes large enough to achieve the geometric limit, can be defined as
\bea
\sigma^{th}_{i \chi} &=& \frac{\pi R_{core}^2}{N_i},
\label{eq:thcore}
\eea
where $R_{core}$, is the radius of the NS core and $N_i$ is the total number of the species $i$. 
However, when particles of species $i$ have a non-zero Fermi momentum, the scattering cross section is suppressed by the factor (see Appendix \ref{sec:suppfact}) 
\begin{eqnarray}
   \frac{\sqrt{\langle q_{tr}^2\rangle_\theta}_{core}}{p_{F,i}^{core}},
\end{eqnarray}
where $q_{tr}$ is the momentum transfer, $\langle \rangle_\theta$ denotes  average over angles  and $p_{F,i}^{core}$ is the Fermi momentum of the species $i$ in the NS core. The threshold cross section is thus larger than that defined in eq.~\ref{eq:thcore} by a factor of $p_{F,i}^{core}/\sqrt{\langle q_{tr}^2\rangle_\theta}_{core}$.

Assuming that DM thermalizes then, after reaching the steady state, the energy contribution of each captured DM particle can be taken as the total initial energy $m_\chi(1/\sqrt{B}-1)$. The DM contribution to the NS luminosity is
\begin{equation}
L^{\infty,th}_{\rm DM} = m_\chi(1/\sqrt{B}-1) C_{\star} B^2= 4\pi \sigma_{SB} R_{core}^2 \left(T_{kin}^{\infty,th}\right)^4, 
\label{eq:lum}
\end{equation}
where $\sigma_{SB}$ is the Stefan-Boltzmann constant and $T^\infty=\sqrt{B}T$ is the temperature measured at large distance from the NS. Using eq.~\ref{eq:Cstar} and eq.~\ref{eq:lum}, we obtain~\cite{Baryakhtar:2017dbj,Bell:2018pkk}
\begin{eqnarray}
T_{kin}^{\infty,th} &=& \left[f \frac{\rho_\chi (1-B)B}{4\sigma_{SB}v_\star }\left(\frac{1}{\sqrt{B}}-1\right)\erf\left(\sqrt{\frac{3}{2}}\frac{v_\star}{v_d}\right)\right]^{1/4}\nonumber \\
&=& 1700 K f^{1/4} \left(\frac{\rho_\chi}{0.4\GeV \cm^{-3}}\right)^{1/4}F\left(\frac{v_\star}{230\km \s^{-1}}\right), 
\label{eq:Tkin}
\end{eqnarray}
where $\sigma_{i \chi}$ is the DM scattering cross section off a particle of species $i$ and $C$ is the NS capture rate,
\begin{equation}
F(x) = \left[\frac{\erf(x)}{x \erf(1)}\right]^{1/4}, 
\end{equation}
and $f \in[0,1]$ is the fraction of incident DM particles captured by the NS, which can be estimated as
\be
f = \frac{C}{C_{\star}} \sim {\rm MIN}\left[\frac{\sigma_{ i \chi}}{\sigma_{th}},1\right].\label{eq:capprox}
\ee
Note from eq.~\ref{eq:Tkin} that a NS blackbody temperature of $T_{kin}^{\infty,th}\simeq1700$ K is expected in the case of maximal DM capture. 
Moreover, in the absence of another heating mechanism, this temperature will lead to radiation in the near infra-red, potentially detectable by the forthcoming JWST \cite{Baryakhtar:2017dbj}. 
It is important to remark that $\sigma_{i \chi}=\sigma_{th}$ maximises the kinetic heating; any larger cross section would produce the same effect as $\sigma_{i \chi}=\sigma_{th}$. For cross sections below $\sigma_{th}$, the capture rate is reduced as $C\propto \sigma_{i \chi}$ and the kinetic heating temperature decreases as  $f^{1/4}$. 

For DM-electron scattering, the averaged momentum transfer for the initial scattering interaction (capture process) is given by 
\begin{equation}
\bar{q} = \sqrt{{\langle q_{tr}^2\rangle_\theta}_{core}} = \sqrt{2 m_e {\langle E_R\rangle_\theta}_{core}},
\end{equation}
where $E_R$ is the energy transfer given by~\cite{Bell:2018pkk}
\bea
E_R &=& \frac{(1-B)m_\chi \mu}{B+2\sqrt{B}\mu+B\mu^2}\left(1-\cos\theta_{cm}\right), 
\label{eq:recoilenel} 
\eea
\begin{equation}
\mu = \frac{m_\chi}{m_e}, 
\end{equation}
and $\theta_{cm}$ is the scattering angle in the centre of mass frame. 

For $m_\chi\gg m_e$, we have
\begin{equation}
\bar{q} \simeq m_e \sqrt{2 \left(\frac{1-B}{B}\right)} \sim 0.9 \MeV .
\end{equation}
For typical nuclear densities at the NS core, $\langle p_{F,e}(r) \rangle = 145.64 \MeV $.
Then, $\frac{\bar{q}}{p_{F,core}} < 1$ for 1 MeV $ \leq m_\chi \leq $ 1 TeV and the scattering cross section is always suppressed by this factor.
The corresponding cross section for muons, on the other hand, is suppressed for $m_\chi \lesssim 100$ MeV.

\begin{figure}[t] 
\centering
\includegraphics[width=0.65\textwidth]{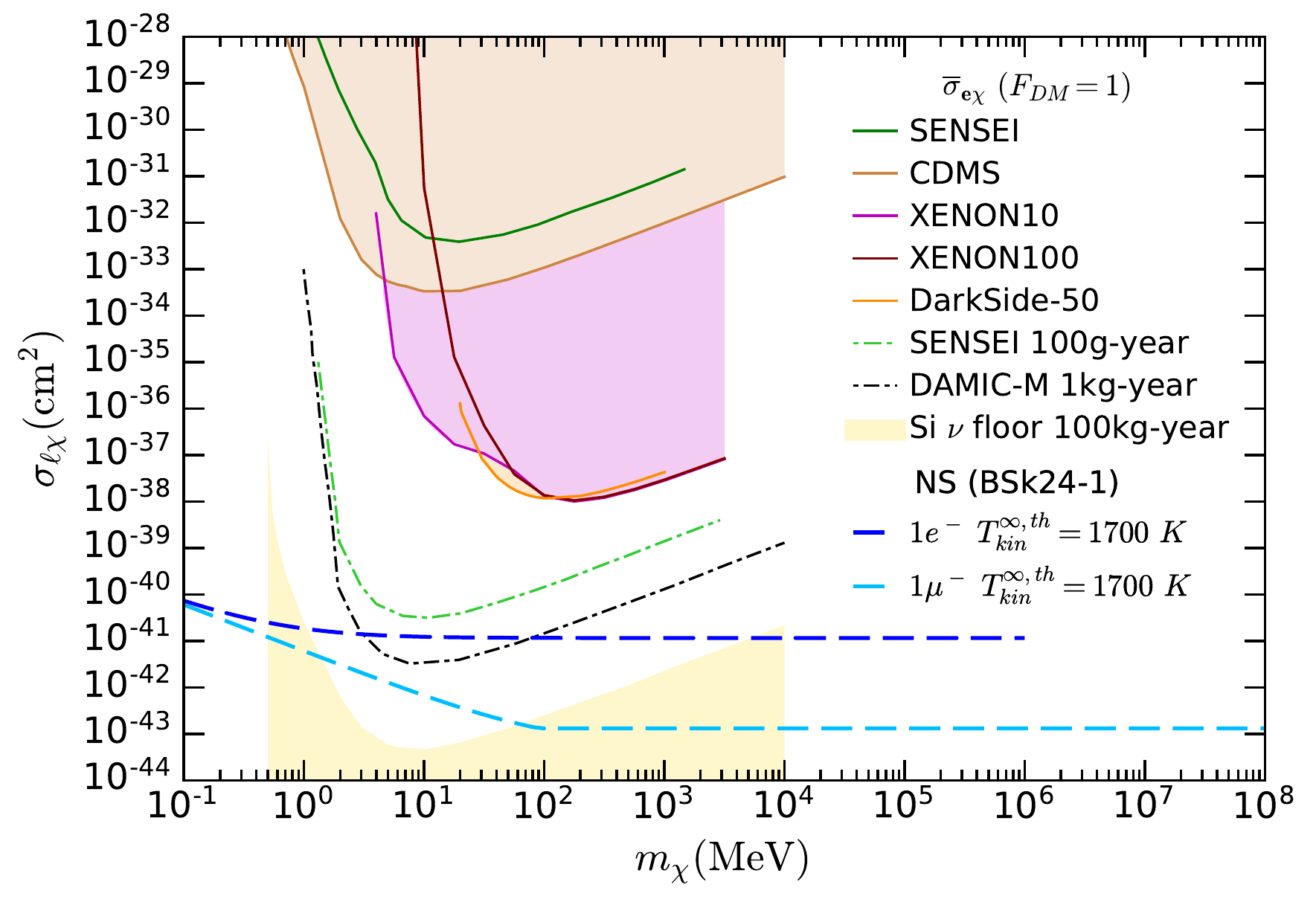}
\caption{NS kinetic heating sensitivity to the DM-electron (blue) and DM-muon (light blue) elastic scattering cross section, for a NS with EoS BSk24-1 and kinetic heating temperature of 1700~K. DD electron recoil limits for heavy mediators from SENSEI, CDMS, XENON10, XENON100 and DarkSide-50 (solid lines),  projected sensitivities from SENSEI and DAMIC-M (dot-dashed lines), and the neutrino background for a silicon target with 100 kg-year exposure (shaded light yellow) are also shown. }
\label{fig:sigmath}
\end{figure}

In Figure \ref{fig:sigmath}, we show the threshold cross section for DM scattering off a single electron (blue dashed line) and a single muon (green dashed line) for our benchmark NS BSk24-1 whose properties can be found in Table~\ref{tab:eos}.  The threshold cross sections have been  calculated with eq.~\ref{eq:thcore} when there is no Pauli blocking and with eq.~\ref{eq:thcorePauliblock} otherwise.
Electron recoil upper limits for heavy mediators ($F_{DM}=1$, i.e. elastic scattering off free electrons\footnote{Recall that bounds from DD electron recoil experiments can be calculated in a model-independent way,  following the parametrization given in refs.~\cite{Essig:2011nj,Essig:2015cda}, i.e. in terms of  $\overline{\sigma}_e=\frac{\mu_{\chi e}^2}{16\pi m_\chi^2 m_e^2}\left. \overline{{ |{\cal M}_{\chi e}(q)|^2}}\right|_{q^2=\alpha^2m_e^2}$, the non-relativistic DM-electron elastic scattering cross section at a fixed momentum transfer $q=\alpha m_e$. The squared matrix element is given by $\overline{ |{\cal M}_{\chi e}(q)|^2}= \left. \overline{{ |{\cal M}_{\chi e}(q)|^2}}\right|_{q^2=\alpha^2m_e^2} \times |F_{DM}(q)|^2$, where $F_{DM}(q)$ is the DM form factor, and $\mu_{\chi e}$ is the DM-electron reduced mass.}) from the SENSEI~\cite{Crisler:2018gci}, CDMS~\cite{Agnese:2018col}, XENON10~\cite{Essig:2012yx}, XENON100~\cite{Essig:2017kqs} and DarkSide-50~\cite{Agnes:2018oej} direct detection experiments, sensitivity projections from SENSEI and DAMIC-M~\cite{Battaglieri:2017aum}, and the neutrino background for a silicon detector with 100 kg-year exposure \citep{Essig:2018tss} are also shown for comparison.

For $m_\chi \gtrsim 1 \TeV$,  multiple scatterings off electrons are required in order for DM particles to be captured, whereas when DM scatters off muons, only one scattering is needed for  $m_\chi \lesssim 10^{8} \MeV$  (following the approach of ref.~\cite{Bramante:2017xlb}). Notice the significant constraining power of the muon threshold cross section, $\sigma^{th}_{\mu\chi}$, which is more sensitive than that of electrons by two orders of magnitude, despite the fact that muons are less abundant in NSs. This sensitivity surpasses future terrestrial DD electron recoil experiments over the entire dark matter mass range and, for much of the parameter space, by several orders of magnitude. Furthermore, NS kinetic heating from DM-muon scattering may potentially probe cross sections below the neutrino floor, especially for $m_\chi \gtrsim 100 \MeV$. At low DM mass, NS DM-muon and DM-electron scattering may allow us to explore the sub-MeV mass regime that is hidden from terrestrial DD experiments.

\subsection{Uncertainties in the threshold cross section}
\label{sec:eos}

\begin{figure}[t]
    \centering
    \includegraphics[width=0.5\textwidth]{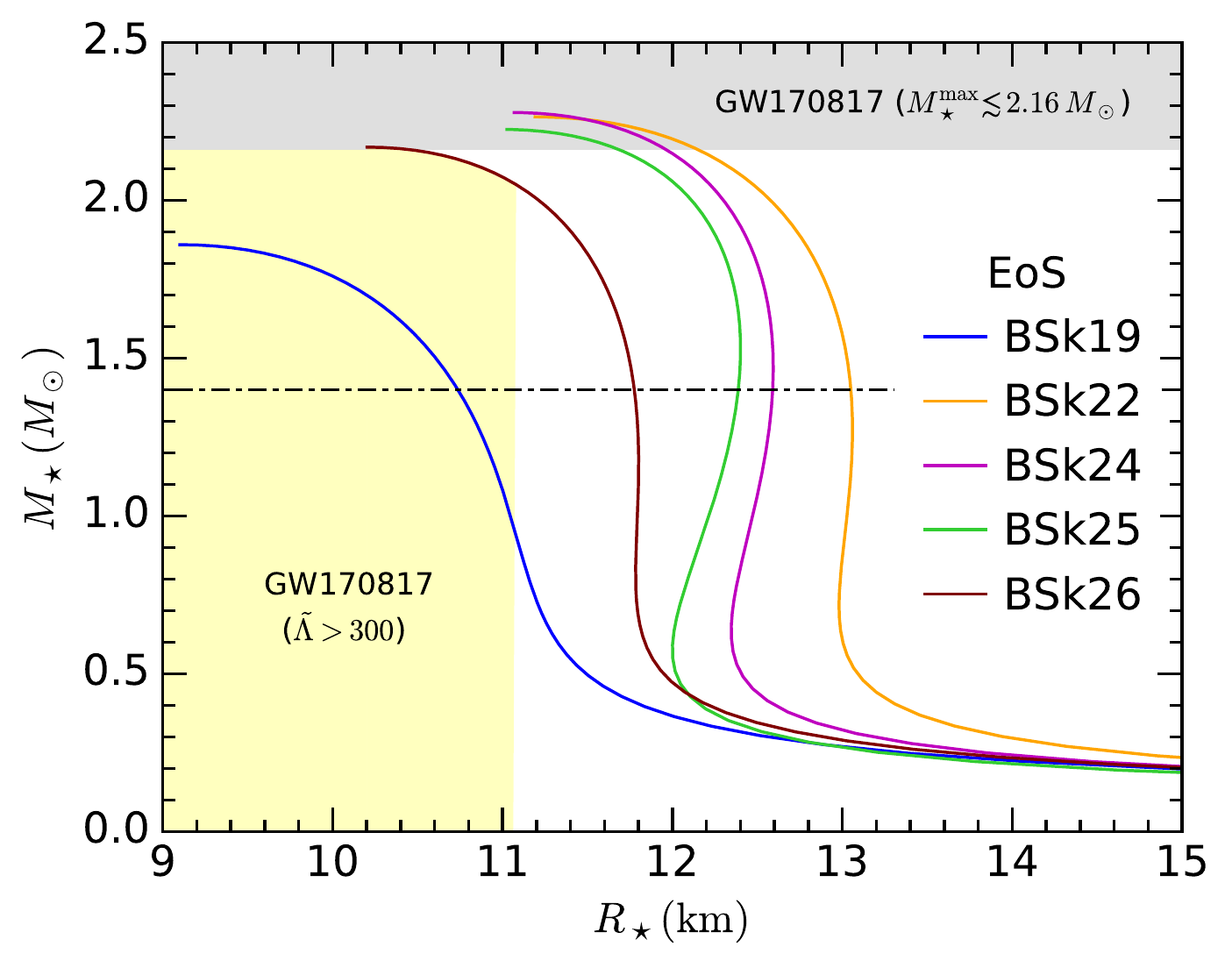}
    \caption{Mass radius relation for the BSk functionals considered here. The grey region denotes the upper bound on the NS mass from the NS binary merger GW170817 \cite{Margalit:2017dij,Shibata:2017xdx,Ruiz:2017due,Rezzolla:2017aly} and the yellow region represents the lower bound on the radius of a NS with $M_\star=1.4M_\odot$ from GW and EM data of GW170817.}
    \label{fig:NS_MR}
\end{figure}

The threshold cross section calculation in the geometric limit for the different species in a NS, i.e. neutron, proton, electrons and muons, relies heavily, not only on the mass and radius of the NS core, but also on the species abundance and its Fermi momentum via Pauli blocking effects. These  microscopic properties exhibit a radial dependence as shown in Figure \ref{fig:YipF}. 

With the aim of assessing the effect that a different EoS and NS configuration choice has  on our results and since we are interested in the kinetic heating of cold NSs by DM capture, we initially considered the unified equations of state for cold non-accreting matter developed by the Brussels-Montreal group \cite{Goriely:2010bm,Pearson:2011,Pearson:2012hz,Goriely:2013}, whose analytical fits, given in refs~\cite{Potekhin:2013qqa,Pearson:2018tkr},  provide us with an excellent tool for evaluating NS microscopic properties without performing NS structure and evolution simulations. 
Moreover, the estimated error of these fits are far below the current observational uncertainties. 
Namely,  we selected the BSk19, BSk22, BSk24, BSk25 and BSk26 functionals, we disregarded BSk20 and BSk21 since their mass radius (MR) relation is very similar to that of BSk26 and BSk24 respectively. 
These functionals take into account only the above stated species as NS constituents, neglecting the possible presence of exotic matter. 

In figure~\ref{fig:NS_MR}, we show the MR relation for the remaining BSk functionals. BSk19 is ruled out by  the observations of the heaviest NSs and by gravitational wave (GW) data from the binary NS merger event GW170817  and its respective electromagnetic (EM) counterpart~\cite{TheLIGOScientific:2017qsa,Abbott:2018wiz,Monitor:2017mdv}, which requires the tidal parameter $\tilde{\Lambda}>300$~\cite{Radice:2018ozg}. This lower bound can be translated into a lower limit for the radius of a NS with $M_\star=1.4M_\odot$ (see yellow shaded region). 
On the other side, the compatibility of the direct Urca process presence in NSs with observations disfavours BSk22 and BSk26 \cite{Pearson:2018tkr}. 
Then, we will taking into account only BS24 and BSk25, introduced in ref.~\cite{Goriely:2013}, 
 which also give the better fits to observational data \cite{Pearson:2018tkr}. 
For every functional, we have considered two possible scenarios, featuring low and large masses. The maximum mass is restricted to  $M_\star\lesssim2.16M_\odot$ by the GW170817 limit~\cite{Margalit:2017dij,Shibata:2017xdx,Ruiz:2017due,Rezzolla:2017aly}. 
Regarding the low mass configuration for BSk24 and Bsk25, we have chosen masses below the direct Urca threshold. 
In Table~\ref{tab:eos}, we summarise the main properties of the aforementioned NS models, where $\langle \rangle $ denotes an average over the volume. 

\begin{table}[tb]
\centering
\begin{tabular}{|l|c|c|c|c|}
\hline
\bf EoS & \bf BSk24-1 & \bf BSk24-2 & \bf BSk25-1 & \bf BSk25-2 \\ \hline
$\rho_c$ $[\rm{g \, cm^{-3}}]$ & $7.76 \times 10^{14}$ & $1.42 \times 10^{15}$ & $7.46 \times 10^{14}$ & $1.58 \times 10^{15}$  \\
$M_\star$ $[M_\odot]$ & 1.500 & 2.160 &  1.400 & 2.160 \\
$R_\star$ [km] & 12.593 & 11.965 & 12.387 & 11.689 \\\hline
\multicolumn{5}{|c|}{\bf NS core} \\\hline
$M_{\rm core}$  $[M_\odot]$ & 1.483 & 2.152 & 1.383 & 2.154 \\
$R_{\rm core}$ [km] & 11.643 & 11.519 & 11.389 & 11.279\\
$\left<Y_n(r)\right>$ & 92.68 \% & 88.86 \% & 93.69 \% & 88.61 \%\\
$\left<Y_p(r)\right>$ & 7.32 \% & 11.14 \% & 6.31 \% & 11.39 \% \\
$\left<Y_e(r)\right>$ & 5.46 \% & 7.25 \% & 4.86 \% & 7.30 \%\\
$\left<Y_\mu(r)\right>$ & 1.85 \% & 3.89 \% & 1.44 \% & 4.09 \%\\
$\left<p_{F,n}(r)\right>$ [MeV] & 372.56 &  410.48 & 374.80 & 417.89\\
$\left<p_{F,p}(r)\right>$ [MeV] & 160.23 & 205.71 & 152.79 & 210.08\\
$\left<p_{F,e}(r)\right>$ [MeV]& 145.64 & 179.19 & 140.31 & 182.58\\
$\left<p_{F,\mu}(r)\right>$ [MeV] & 50.38 &  75.26 & 45.66 &  77.42 \\\hline 
\end{tabular} 
\caption{Benchmark NSs, for two different equations of state (EoS) for cold non-accreting neutron stars with Brussels–-Montreal functionals BSk24 and BSk25 \cite{Pearson:2018tkr} and two mass regimes determined by the central mass-energy density $\rho_c$. The microscopic properties of the core, species abundances and Fermi momentum, have been averaged over the  volume.  
}
\label{tab:eos}
\end{table}

\begin{figure}[t] 
\centering
\includegraphics[width=0.7\textwidth]{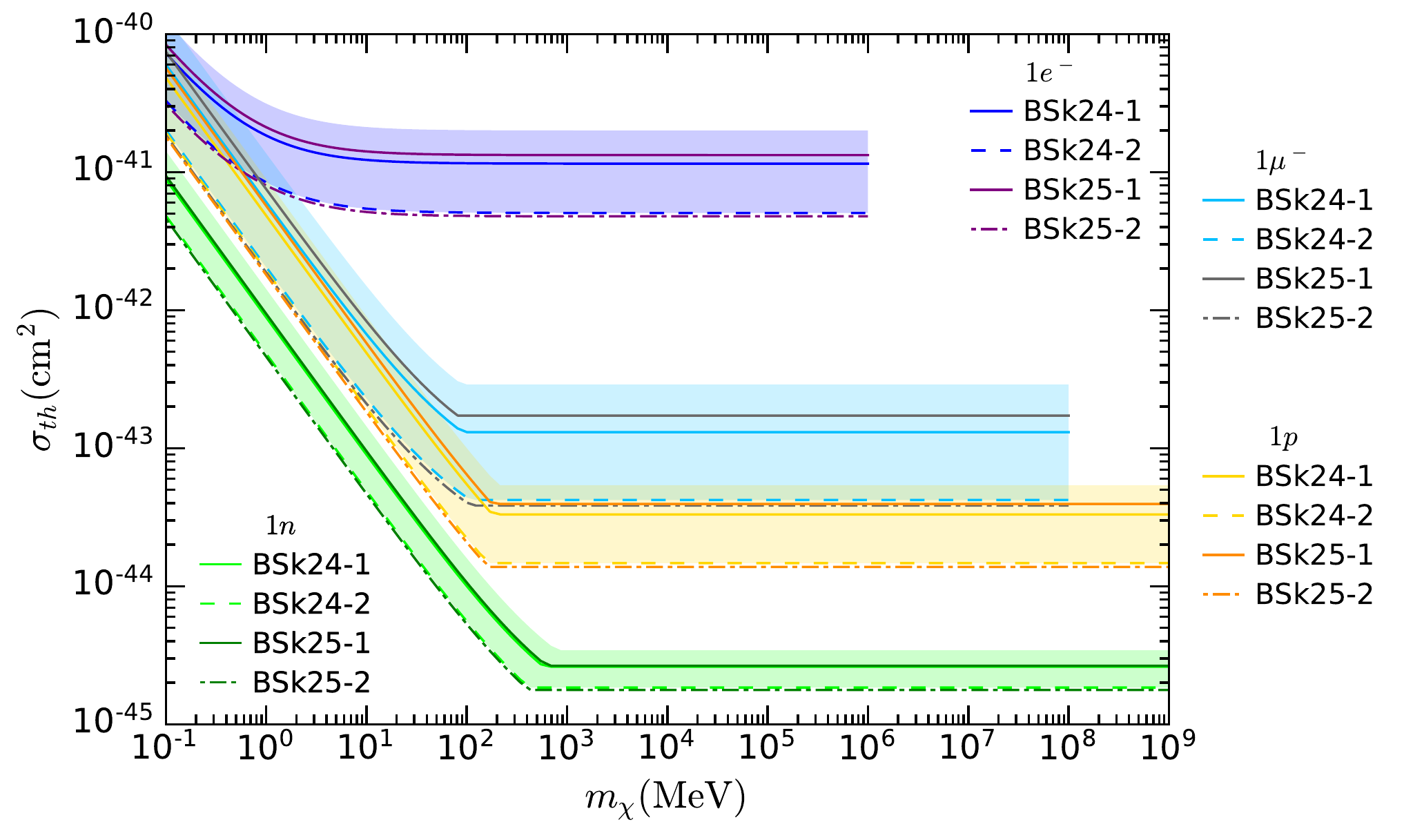}
\caption{NS threshold cross section for DM scattering off electrons (blue and purple lines),  muons (light blue and grey lines), protons (yellow and orange lines) and neutrons (light and dark green lines) for the benchmark models in Table~\ref{tab:eos}. The shaded regions correspond to the variation in $\sigma_{th}$ from larger to smaller cross sections for the BSk24  functional with NS masses in the 1 --  2.16 $M_\odot$ range.
}
\label{fig:sigmath_eos}
\end{figure}

As previously stated, 
we have restricted our analysis to the NS core, since it contains $\sim99\%$ of the mass of the star. This is an appropriate and convenient choice since our aim is to determine the NS kinetic heating sensitivity to leptophilic DM scattering cross sections, and the lepton density in the crust is much lower than in the core. Using the NS core properties in Table~\ref{tab:eos} together with either eq.~\ref{eq:thcorePauliblock} when there is Pauli blocking,  or with eq.~\ref{eq:thcore} otherwise, we have computed the threshold cross section for every benchmark model and every species in a NS. Our results are depicted in Figure~\ref{fig:sigmath_eos}. 
As expected from Figure~\ref{fig:NS_MR},  changing the equation of state from BSk24 to Bsk25 has almost no impact on the threshold cross section for large mass configurations: compare  blue (BSk24-2) with purple dashed lines (BSk25-2) for DM scattering off electrons, light blue (BSk24-2) with grey dashed lines (BSk25-2) for muons, yellow  (BSk24-2) with orange dashed lines (BSk25-2)  for protons and light green (BSk24-2) with dark green (BSk25-2) dashed lines for neutrons. For the low mass configurations, the difference is more noticeable due to the mass separation between BSk24-1 ($M_\star=1.5M_\odot$) and BSk25-1 ($M_\star=1.4M_\odot$), yet still very small. In fact, the difference between two mass configurations of the functionals BSk24 and BSk25 is insignificant compared to the difference that occurs if we move from a low mass configuration (large $\sigma_{th}$) such as  $1M_\odot$ to a large mass regime (small $\sigma_{th}$) along the same functional (see shaded regions in Figure~\ref{fig:sigmath_eos}). Remarkably, the combined effect of choosing a different EoS and mass configuration translates to less than one order magnitude change in the threshold cross section for muon, electron or proton scattering. The effect is even less pronounced for neutrons since they are the dominant species in NSs. In light of the foregoing, we have selected the model BSk24-1 as our benchmark NS throughout this paper, since a low mass configuration allows us to derive more conservative bounds on the cutoff scale of the fermionic DM operators than a NS with larger mass. We have chosen EoS BSk24 over BSk25, since the former gives slightly better NS mass fits than the latter \cite{Pearson:2018tkr}.

  \begin{table}
  \small
\centering
{\renewcommand{\arraystretch}{1.3}
\begin{tabular}{ | c | c | c | c |}
  \hline                        
  Name & Operator & Coupling $G$ & $\frac{d\sigma}{d\cos\theta}(s,t)$   \\   \hline
  L1 & $\bar\chi  \chi\;\bar \ell  \ell $ & ${y_\ell}/{\Lambda^2}$  & $\frac{y_\ell^2}{\Lambda^4} \frac{\left(4 m_{\chi }^2-t\right) \left(4 m_{\chi }^2-\mu ^2
   t\right)}{32 \pi  \mu ^2 s}$ \\  \hline
  L2 & $\bar\chi \gamma^5 \chi\;\bar \ell \ell $ & $i{y_\ell}/{\Lambda^2}$  & $\frac{y_\ell^2}{\Lambda^4} \frac{t \left(\mu ^2 t-4 m_{\chi }^2\right)}{32 \pi  \mu ^2 s}$ \\  \hline
  L3 & $\bar\chi \chi\;\bar \ell \gamma^5  \ell $ &  $i{y_\ell}/{\Lambda^2}$ & $\frac{y_\ell^2}{\Lambda^4} \frac{ t \left(t-4 m_{\chi }^2\right)}{32 \pi  s}$ \\  \hline
  L4 & $\bar\chi \gamma^5 \chi\; \bar \ell \gamma^5 \ell $ & ${y_\ell}/{\Lambda^2}$ & $\frac{y_\ell^2}{\Lambda^4} \frac{ t^2}{32 \pi s}$ \\  \hline
  L5 & $\bar \chi \gamma_\mu \chi\; \bar \ell \gamma^\mu \ell$ & ${1}/{\Lambda^2}$ & $\frac{1}{\Lambda^4} \frac{2 \left(\mu ^2+1\right)^2 m_{\chi }^4-4 \left(\mu ^2+1\right) \mu ^2 s m_{\chi }^2+\mu ^4 \left(2 s^2+2 s t+t^2\right)}{16
   \pi  \mu^4 s}$ \\  \hline
  L6 & $\bar\chi \gamma_\mu \gamma^5 \chi\; \bar  \ell \gamma^\mu \ell $ & ${1}/{\Lambda^2}$ & $\frac{1}{\Lambda^4} \frac{2 \left(\mu ^2-1\right)^2 m_{\chi }^4-4 \mu ^2 m_{\chi }^2 \left(\mu ^2 s+s+\mu ^2 t\right)+\mu ^4 \left(2 s^2+2 s
   t+t^2\right)}{16 \pi  \mu^4 s}$  \\  \hline
  L7 & $\bar \chi \gamma_\mu  \chi\; \bar \ell \gamma^\mu\gamma^5  \ell$ & ${1}/{\Lambda^2}$  & $\frac{1}{\Lambda^4} \frac{2 \left(\mu ^2-1\right)^2 m_{\chi }^4-4 \mu ^2 m_{\chi }^2 \left(\mu ^2 s+s+t\right)+\mu ^4 \left(2 s^2+2 s t+t^2\right)}{16
   \pi  \mu^4 s}$ \\  \hline
  L8 & $\bar \chi \gamma_\mu \gamma^5 \chi\; \bar \ell \gamma^\mu \gamma^5 \ell $ & ${1}/{\Lambda^2}$ &  $\frac{1}{\Lambda^4} \frac{2 \left(\mu ^4+10 \mu ^2+1\right) m_{\chi }^4-4 \left(\mu ^2+1\right) \mu ^2
   m_{\chi }^2 (s+t)+\mu ^4 \left(2 s^2+2 s t+t^2\right)}{16 \pi  \mu ^4 s}$ \\  \hline
  L9 & $\bar \chi \sigma_{\mu\nu} \chi\; \bar \ell \sigma^{\mu\nu} \ell $ & ${1}/{\Lambda^2}$ & $\frac{1 }{\Lambda^4} \frac{4 \left(\mu ^4+4 \mu ^2+1\right) m_{\chi }^4-2 \left(\mu ^2+1\right) \mu ^2 m_{\chi
   }^2 (4 s+t)+\mu ^4 (2 s+t)^2}{4 \pi  \mu ^4 s}$  \\  \hline
 L10 & $\bar \chi \sigma_{\mu\nu} \gamma^5\chi\; \bar \ell \sigma^{\mu\nu} \ell \;$  & ${i}/{\Lambda^2}$  & $\frac{1}{\Lambda^4} \frac{4 \left(\mu ^2-1\right)^2 m_{\chi }^4-2 \left(\mu ^2+1\right) \mu ^2 m_{\chi }^2 (4 s+t)+\mu ^4 (2 s+t)^2}{4 \pi  \mu^4 s}$ \\  \hline
\end{tabular}}
\caption{Operators and differential cross sections for Dirac DM scattering off leptons. The effective couplings for each operator are given as a function of the lepton Yukawa coupling, $y_\ell$, and the cutoff scale, $\Lambda$. The fourth column is the differential cross section at high energy as a function of the Mandelstam variables $s$ and $t$. 
\label{tab:operators}}
\end{table}

\section{DM-lepton scattering cross sections}
\label{sec:leptons}
We have considered the case of fermionic DM that couples directly only to leptons, and followed a model independent approach. The four-fermion interactions of DM with SM leptons are then described by the full list of dimension 6 Effective Field Theory (EFT) operators, as classified in ref.~\citep{Goodman:2010ku} (for DM interactions with SM quarks). These operators, and the corresponding differential cross sections for elastic DM-lepton scattering in terms of the Mandelstam variables $s$ and $t$, are given in Table~\ref{tab:operators}, where $\mu=\frac{m_\chi}{m_\ell}$. More details about the cross section computation for leptons and nucleons can be found in Appendix~\ref{sec:xsec}. 
The gravitational pull of a NS will accelerate DM particles to velocities comparable to the speed of light, overcoming velocity and momentum suppression, which is the main advantage of NSs over Earth-based experiments. Therefore, we have calculated the full high energy form of the cross sections (to be contrasted with the non-relativistic approximations that are used for DD calculations\footnote{The differential cross sections in the non-relativistic limit can be obtained by expanding the Mandelstam variables in powers of the relative speed, $w$, and the momentum transfer, $q_{tr}$, and keeping only the largest non-zero term. A list of these expressions for scattering off nucleons can be found in Table~2 of ref.~\cite{Bell:2018pkk}.}).

We have assumed that NSs are made only of neutrons, protons, electrons and muons, and  that a single interaction with one of the aforementioned species is responsible for the DM capture. Hence, we have restricted our analysis to the region of DM mass where the momentum transfer in a collision with any one of these four species is sufficient for capture, i.e. $1\MeV \leq m_\chi \leq 10^6\MeV$. In the absence of significant additional heating mechanisms, the NS equilibrium temperature will be set by the capture process.

\section{Loop induced couplings to quarks}
\label{sec:loops}

\begin{figure}[t] 
\centering
\includegraphics[width=0.45\textwidth]{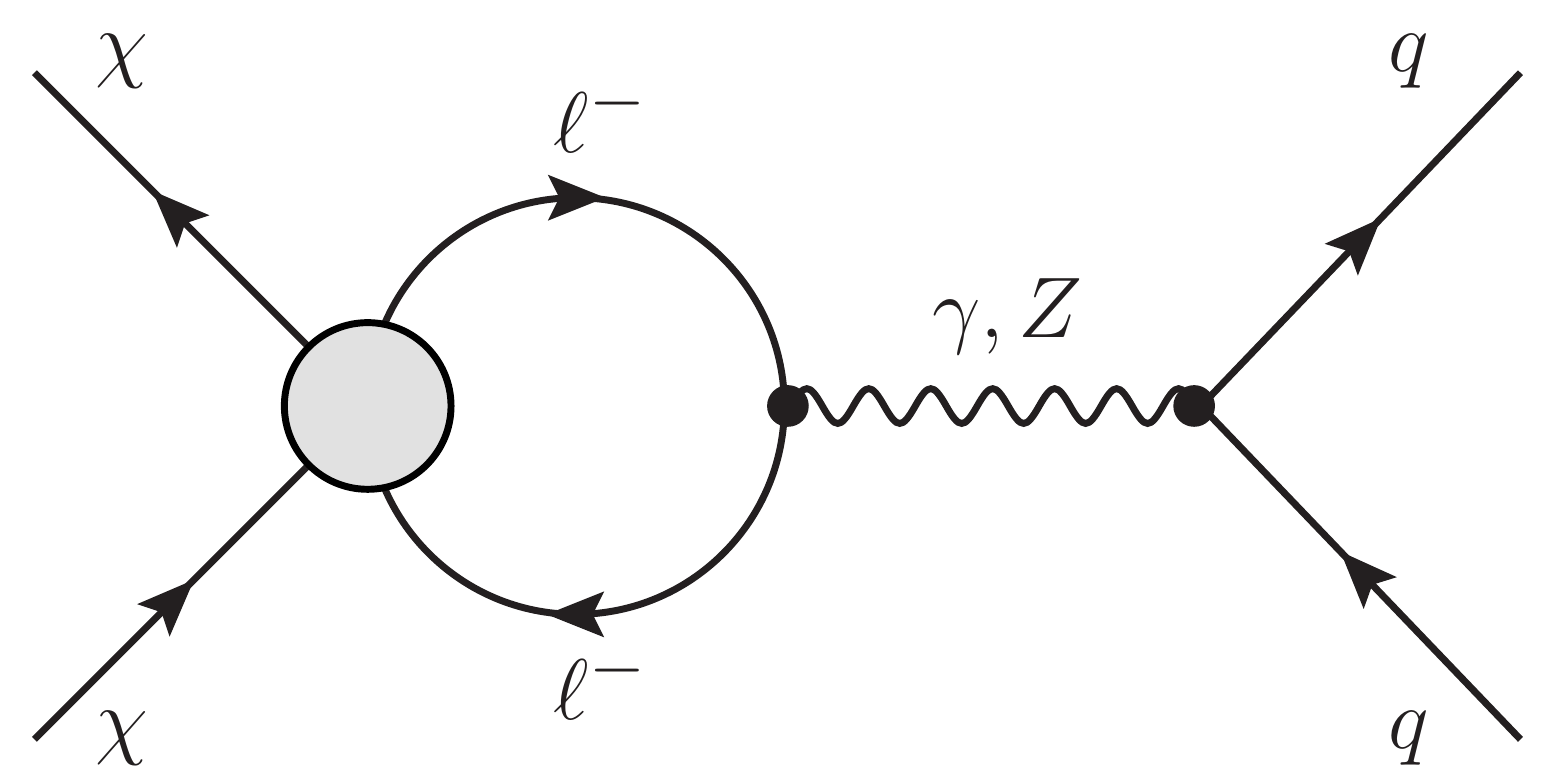}
\caption{One loop DM-quark interaction induced by SM photon and Z boson exchange. }
\label{fig:oneloop}
\end{figure}

\begin{figure}[t] 
\centering
\includegraphics[width=0.45\textwidth]{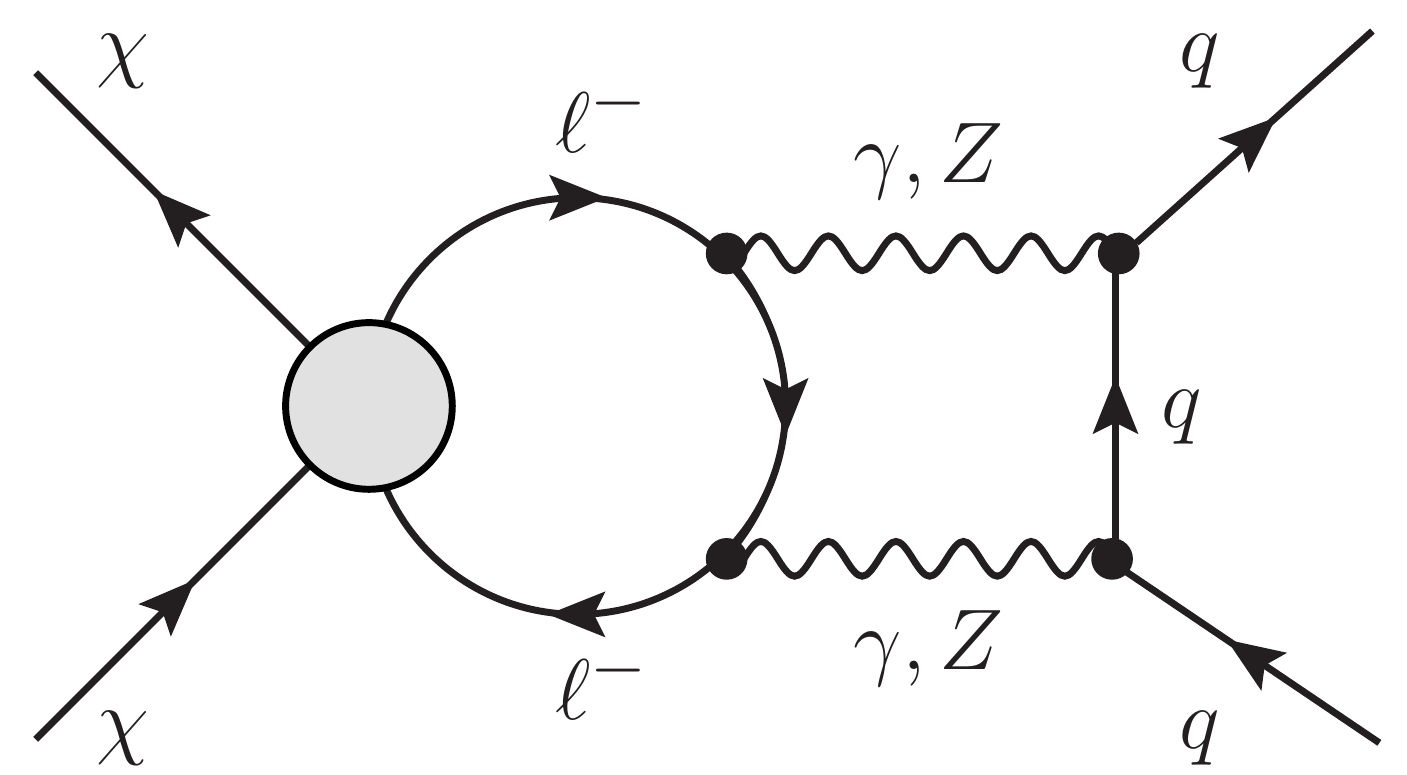}
\hspace{1cm}
\includegraphics[width=0.45\textwidth]{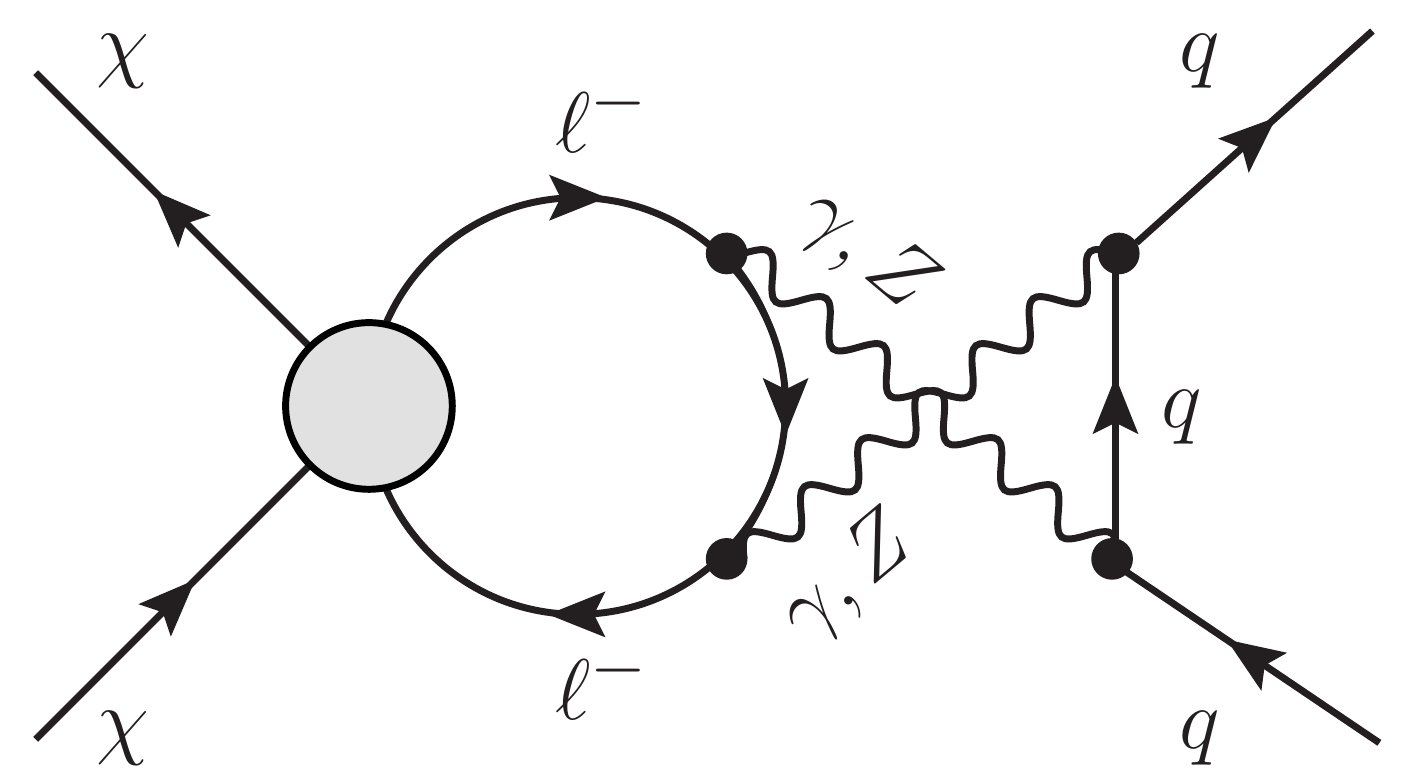}
\caption{Two loop induced DM-quark interaction mediated by SM photon and Z boson exchange.}
\label{fig:twoloops}
\end{figure}

As mentioned before, we have assumed the DM is fermionic and coupled only to leptons. However, couplings to quarks will inevitably be induced at loop level. In Table~\ref{tab:couplings} we indicate where contributions to the DM-quark operators are induced at loop level by the DM-lepton operators, and at what loop order. The relevant loop diagrams are shown in Fig.~\ref{fig:oneloop} and Fig.~\ref{fig:twoloops}, and involve the exchange of either a photon or $Z$, with the latter suppressed by powers of either $q_{tr}^2/M_Z^2$ or $m_l^2/M_Z^2$. (There are similar diagrams involving the exchange of a Higgs, which we neglect as they would be suppressed by powers of $1/M_H^2$ and by the small lepton Yukawa couplings.)
In the case of scalar operators, where loop induced couplings arise only at 2 loops, we therefore consider only contributions from $\gamma$ loops. 

When calculating loop contributions in an EFT, it is important to consider the validity of the approximations employed. Specifically, (i) we may assume that the energy scale of the UV physics that is integrated out to obtain the EFT description is much more massive than the energy or mass of any particle appearing in the loop process. If this is true, any loop diagram in the UV theory is well approximated by the one obtained by first integrating out heavy particles to obtain EFT operators, and then calculating the loop using this EFT. Alternatively, (ii) a loop contribution in the UV theory is also well approximated by a loop calculation in EFT if integrating out the heavy particles factorizes from the loop. 

Now, given that our aim is to explore a wide range of DM masses, up to $\TeV$ scale, option (i) is not guaranteed to be satisfied. Option (ii), instead, could be easily achievable when operators L1-L8 induce D1-D8. As an example, consider a simple UV model in which an s-channel mediator couples to both a lepton bilinear and a DM bilinear. Depending on the spin of the mediator and the nature of its couplings, the EFT description would include one, or a mixture, of the operators L1-L8. Due to the s-channel structure of the couplings, Feynman diagrams like that of Fig.~\ref{fig:oneloop} would factorize into a part containing the new heavy propagator and another containing the lepton loop. In other words, the heavy (integrated out) propagator would not appear as an internal line in the loop.

If we instead consider operators L9 and L10, we encounter additional subtleties. These operators usually appear as linear combinations in models with a t-channel propagator. While they can be Fierz-transformed to to-channel form, the mediator would now appear as an internal line in the loop. Therefore, integrating out the mediator no longer commutes with the loop evaluation, unless there is a clear hierarchy in the masses of the particles participating.
An additional issue, when considering loops involving L9 and L10, is that they generate long-range interactions between DM particles and ordinary matter. 
Hence the evaluation of loop induced DM-quark interactions is best performed in a UV theory, and thus we will not evaluate DM-quark couplings for the L9 and L10 lepton EFT operators.

The dominant loop contributions induced by L1-L8 are calculated below. We note that 
operators L2 and L4 do not induce couplings to quarks at either 1 or 2 loops, while operators D2 and D4 do not receive contributions at 1 or 2 loops.

\begin{table}[tb]\centering
\begin{tabular}{|c|c|c|}
\hline
Operator & Coupling & Induced by\\ \hline
D1 & 2 loop ($\gamma$, Z) & L1\\\hline
D2 & - & -\\\hline
D3 & 2 loop ($\gamma$, Z) & L3\\\hline
D4 & - & -\\\hline
D5 & 1 loop ($\gamma$) & L5 \\
   & 1 loop (Z) & L5, L7\\\hline
D6 & 1 loop ($\gamma$) & L6\\
   & 1 loop (Z) & L6, L8\\\hline
D7 & 1 loop (Z) & L5, L7\\\hline
D8 &  1 loop (Z) & L6, L8\\\hline
\end{tabular}
\caption{DM-quarks couplings generated at loop level by lepton interactions. 
We note that the $Z$ contributions are suppressed by powers of either $q_{tr}^2/M_Z^2$ or $m_l^2/M_Z^2$. (Also note that the momentum suppression factors differ from those for standard momentum suppressed DM-quark scattering, which are proportional to powers of $q_{tr}^2/m_\chi^2$.)}\label{tab:couplings}
\end{table}

\subsection{One loop $\gamma$-induced couplings}

The one loop contributions are generated by mixing of the dark mediator with the photon and Z bosons of the SM. They are all proportional to the divergent part of \citep{Kopp:2009et}

\begin{equation}
    \int\frac{d^4 k}{(2\pi)^4} \frac{Tr[\Gamma_i (\slashed{k}+m_\ell)\Gamma_j^{\gamma,Z}(\slashed{k}+\slashed{q}+m_\ell)]}{(k^2-m_\ell^2)((k+q)^2-m_\ell^2)}, 
\end{equation}
where $\Gamma$ denotes different Lorentz structures. 

Considering only photon exchange, for the case of vector couplings to SM leptons (L5, L6), we obtain
contributions to the operators
\begin{eqnarray}
    && G_q^5\bar{\chi}\gamma_\mu\chi \bar{q}\gamma^\mu q,
    \label{eq:G5}\\
    && G_q^6\bar{\chi}\gamma_\mu\gamma_5\chi \bar{q}\gamma^\mu q,
    \label{eq:G6}
\end{eqnarray}
with induced couplings given by 
\begin{eqnarray}
G_q^{5} &=& Q_q \frac{\alphaem}{3\pi} \sum_l G_\ell^{5} Q_\ell \log\frac{m_\ell^2}{\lambda^2},\\
G_q^{6} &=& Q_q \frac{\alphaem}{3\pi} \sum_l G_\ell^{6} Q_\ell \log\frac{m_\ell^2}{\lambda^2},
\end{eqnarray}
where $\lambda$ is the center of mass energy, $G^{5,6}_\ell$ are the L5 and L6 couplings to leptons, and $Q_\ell$ and $Q_q$ are the lepton and quark electromagnetic charge respectively. Note, however, that in some cases the dependence on $\lambda$ cancels out. For example, in a $L_e-L_\mu$ model \citep{Duan:2017qwj}, where $G_e^5=-G_\mu^5=G$, we obtain
\begin{equation}
    G_q^5 = G Q_q \frac{\alphaem}{3\pi}\log\frac{m_e^2}{m_\mu^2}.
\end{equation}

\subsection{One loop Z-induced couplings}

The L5 and L6 lepton operators induce mixtures of the D5 and D6 quark operators via 1-loop diagrams involving Z-exchange. Likewise, the  L7 and L8 operators induce contributions to D7 and D8. These contributions are all suppressed, either by $q_{tr}^2/M_Z^2$ or $m_l^2/M_Z^2$.  The expressions for the induced quark couplings are given in terms of the lepton couplings as:
\begin{eqnarray}
     G_q^5 &=& G_\ell^5 c_v^q f_v + G_\ell^7 c_v^q f_a,\\
     G_q^6 &=& G_\ell^6 c_v^q f_v + G_\ell^8 c_v^q f_a,\\
     G_q^7 &=& G_\ell^5 c_a^q f_v + G_\ell^7 c_a^q f_a,\\
     G_q^8 &=& G_\ell^6 c_a^q f_v + G_\ell^8 c_a^q f_a,\\ 
     f_v &=& \frac{c_v^\ell}{3\pi} \frac{q_{tr}^2}{M_Z^2} \log\frac{m_\ell^2}{\lambda^2},\\
     f_a &=& \frac{c_a^\ell}{\pi}\left(-\frac{3 m_\ell^2}{M_Z^2} + \frac{9 m_\ell^2+ M_Z^2}{3 M_Z^2} \frac{q_{tr}^2}{M_Z^2}\right) \log\frac{m_\ell^2}{\lambda^2},
\end{eqnarray}
where $c_v^f, c_a^f$ are the vector and axial couplings of the fermion $f$ to the $Z$ boson.

\subsection{Two loop $\gamma$-induced couplings}

In the case of scalar interactions, the coupling to quarks arises instead at two loop order and is given by
\begin{equation}
    G_q^{1,3} = Q_q^2 \frac{\alphaem^2}{2\pi^2}\sum_\ell G_\ell^{1,3}  \int d^4j \frac{3m_\ell m_q(4m_\ell^2-j^2)j^2+6m_\ell^3m_q\sqrt{j^2(j^2-4m_\ell^2)}\log\frac{2m_\ell^2-j^2+\sqrt{j^2(j^2-4m_\ell^2)}}{2m_\ell^2}}{j^6(j^2-m_q^2)(j^2-4m_\ell^2)}.
\end{equation}
This can be approximated by the simple expression
\begin{equation}
    G_q^{1,3} = Q_q^2 \frac{\alphaem^2}{2\pi^2}\sum_\ell G_\ell^{1,3} \frac{2m_\ell m_q}{4m_\ell^2-m_q^2}\log \frac{4m_\ell^2}{m_q^2} \le Q_q^2\frac{\alphaem^2}{2\pi^2}\sum_\ell \hat{y}_\ell^S ,
\end{equation}
which is of the same order as the exact result.

\subsection{Effective couplings to nucleons}

In the previous subsections we have calculated the loop induced couplings of DM to quarks. In order to calculate the DM scattering cross section off neutrons and protons, we need to compute the couplings of the nucleon level operators, taking into account the hadronic matrix elements, as outlined in Appendix~\ref{sec:xsec}. 
For operators featuring loop-level quark couplings induced by $\gamma$ exchange, L1, L3, L5, L6, L9 and L10,  the DM-proton effective couplings are dominated by the electromagnetic contribution, while the loop-level Z-induced contribution is subleading. DM-neutron effective couplings, on the other hand, are always dominated by loop-level contributions induced via Z exchange.
Note that for L9 and L10 the non-relativistic DM-proton couplings have previously been calculated in the literature \cite{Kopp:2009et}; We will use those results for estimating the DD bounds from elastic scattering off protons in the following section.

\section{Results}
\label{sec:results}

\begin{figure}[t] 
\centering
\includegraphics[width=\textwidth]{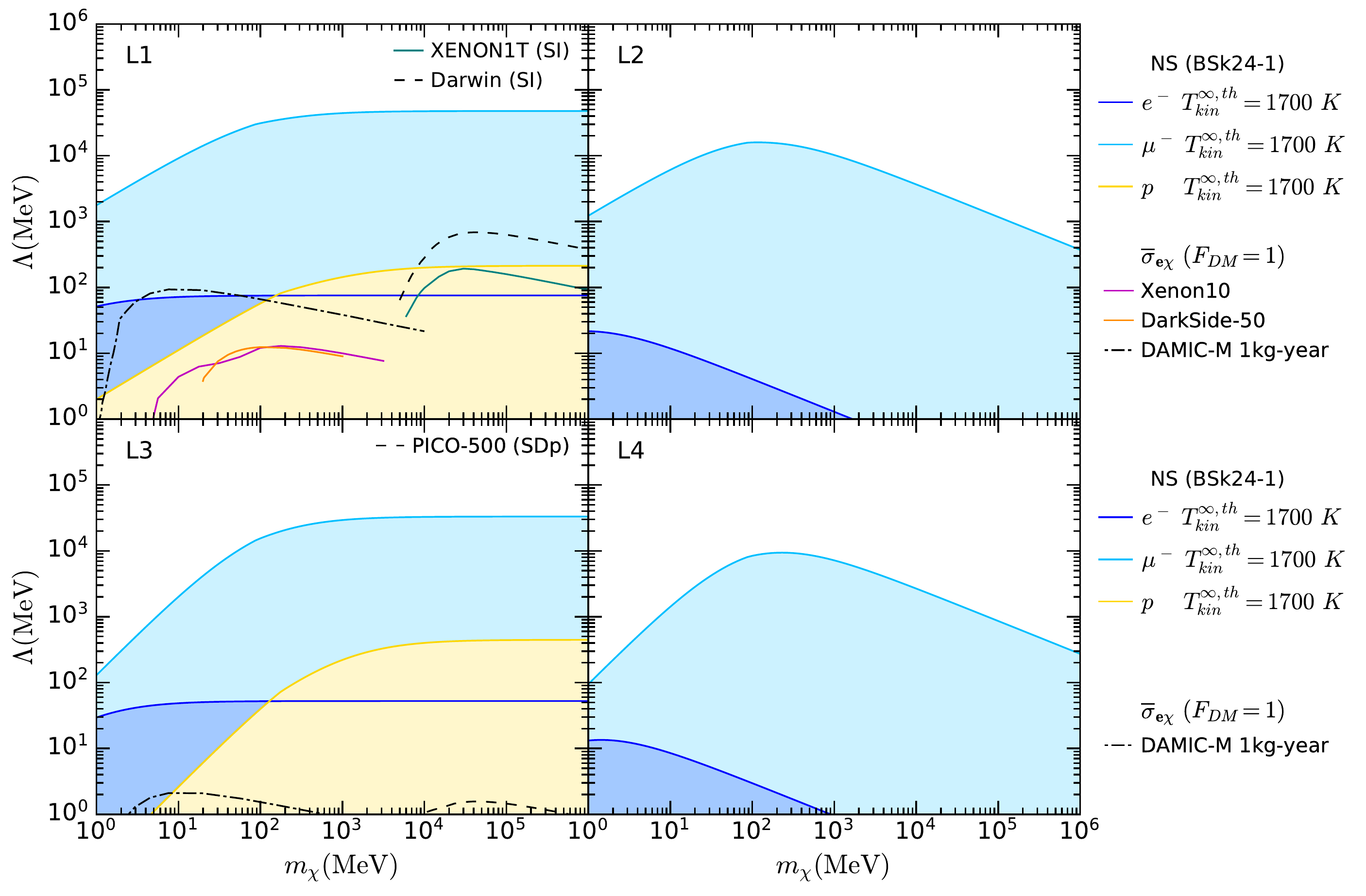}
\caption{Contours of $\sigma=\sigma_{th}$ for leptophilic DM scattering off electrons (blue), muons (light blue), neutrons (green) and protons (yellow), corresponding to $T_{kin}^{\infty,th}=1700 $ K for the operators L1 -- L4 in Table~\ref{tab:operators}. 
Limits from the leading DD electron recoil experiments for heavy mediators are depicted as solid lines, XENON10 (violet) and  DarkSide-50 (orange) and the  projected bounds for DAMIC-M 1kg-year exposure. as a black dot-dashed line.
The solid teal line is the upper limit from XENON1T (SI)  and the dashed lines are the projected bounds for the DARWIN and PICO-500 experiments. }
\label{fig:D1_D4}
\end{figure} 

\begin{figure}[t] 
\centering
\includegraphics[width=\textwidth]{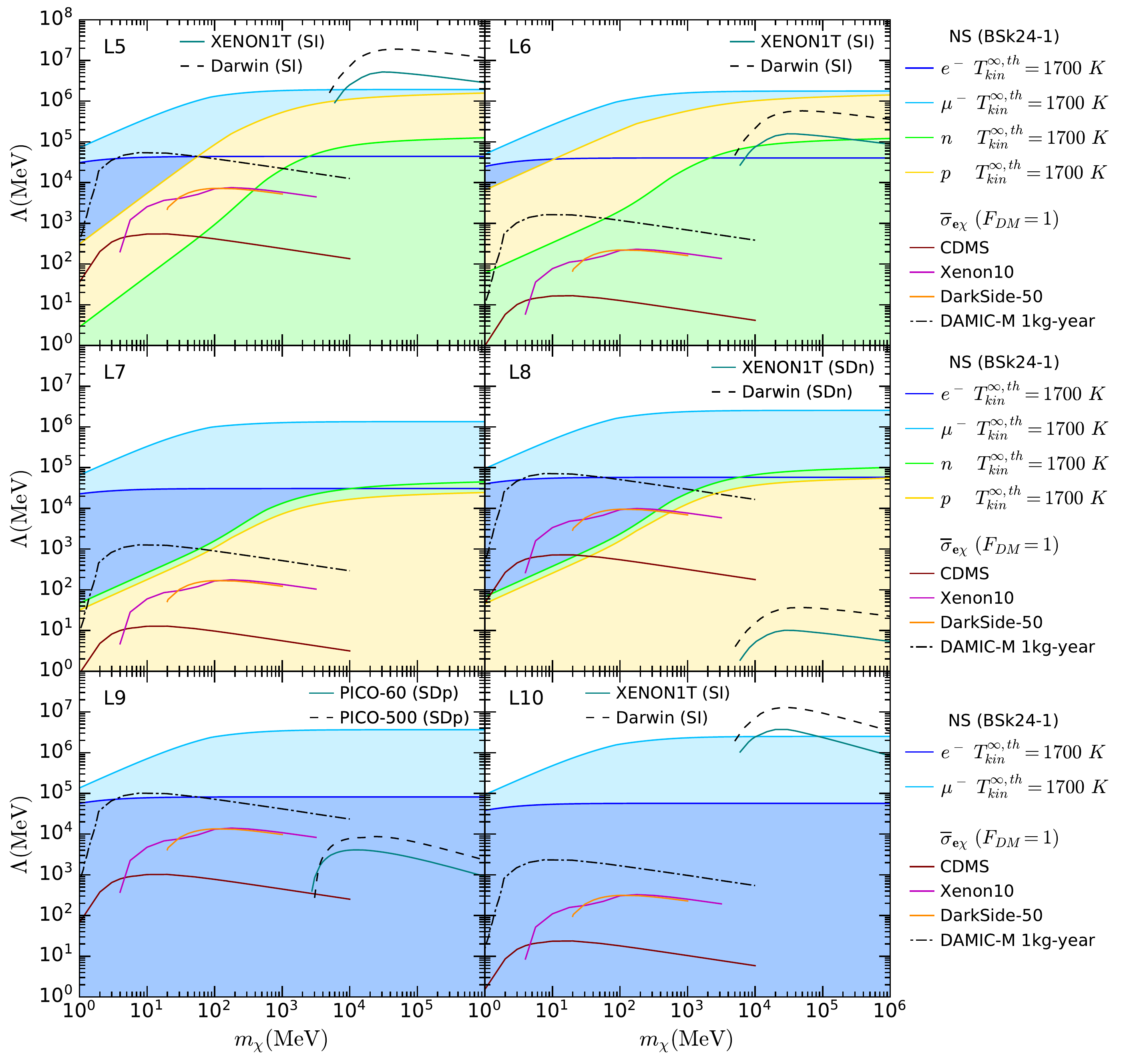}
\caption{Contours of $\sigma=\sigma_{th}$ for leptophilic DM scattering off electrons (blue), muons (light blue), neutrons (green) and protons (yellow), corresponding to $T_{kin}^{\infty,th}=1700 $ K for the operators L5-- L10. 
Limits from the leading DD electron recoil experiments for heavy mediators are depicted as solid lines, CDMS (brown), XENON10 (violet) and  DarkSide-50 (orange) and the  projected bounds for DAMIC-M 1kg-year exposure as black dot-dashed lines.
The solid teal lines are the upper limits from XENON1T (SI  and SD-neutron), PICO-60 (SD-proton) and the dashed lines are the projected bounds for the DARWIN and PICO-500 experiments. }
\label{fig:D5_D10}
\end{figure} 

In Figures~\ref{fig:D1_D4} and \ref{fig:D5_D10}, we present contours of $T_{kin}^{\infty,th}=1700$ K, equivalent to $\sigma=\sigma_{th}$, in the $m_\chi$-$\Lambda$ plane for  DM scattering off electrons, muons, neutrons and protons~\footnote{As mentioned in the previous section, we have not calculated NS upper bounds from DM scattering off neutrons and protons for D9 and D10.}, using our benchmark NS BSk24-1. 
Upper bounds from the leading DD experiments searching for electron recoils ($F_{DM}=1$) and the sensitivity projections for 1kg-year exposure of the DAMIC-M experiment are also shown, as well as upper bounds from the XENON1T (SI)~\cite{Aprile:2018dbl} and (SD neutron)~\cite{Aprile:2019dbj}, PICO-60 (SD proton)~ \cite{Amole:2019fdf} and the future DARWIN~\cite{Aalbers:2016jon} and PICO-500 experiments~\cite{VasquezJauregui:2017} that arise from loop-induced couplings to quarks. For the computation of direct detection bounds, we have employed the non-relativistic limits of the relevant cross sections.

As expected from Figure~\ref{fig:sigmath}, NS kinetic heating via DM-muon scattering provides more stringent upper limits than DM-electron scattering and than terrestrial electron recoil direct detection experiments, for all the operators. This is due to the lower (more constraining) threshold cross section for muons. 
Even though neutrons and protons are more abundant in NSs than leptonic species and therefore their threshold cross sections are lower (the threshold cross section for neutrons is about two orders of magnitude lower than that of muons, see figure~\ref{fig:sigmath_eos}), the DM coupling to their constituents quarks is induced at either one or two loops and hence they are either ${\cal O}(\alphaem)$ or ${\cal O}(\alphaem^2)$ suppressed, respectively. It is worth noting that, nevertheless, the kinetic heating sensitivity for loop-induced DM-nucleon scattering is better than that of tree-level DM-electron scattering, except for low DM mass. 
The slope observed at lower masses for muons, protons and neutrons is due to Pauli blocking suppression (see Figure~\ref{fig:sigmath}). 

For the scalar/pseudoscalar operators L1, L2, L3 and L4 (see figure~\ref{fig:D1_D4}) the difference between the muon and electron sensitivity is particularly pronounced since the couplings of these operators are proportional to the lepton Yukawas. As a result, the bounds on L1 -- L4 are all of the same order of magnitude (aside from the fact that the limits for L2 and L4 become weaker at masses larger than the corresponding lepton mass because their cross sections are suppressed by a factor $1/\mu^2$). Remarkably, the upper limits from DM scattering off electrons in NSs are far more sensitive than current Earth-based electron recoil experiments; Only for L1 do the NS DM-electron scattering limits become comparable to the 1kg-year exposure of the DAMIC-M experiment for sub-GeV DM and outperformed by XENON1T (SI) upper limits for $m_\chi\gtrsim 10\GeV$. For the remaining scalar operators, L2, L3 and L4, bounds from DM capture in NSs with any possible target surpass any current and forthcoming terrestrial DD experiment.

Regarding operators L5 -- L10, (see Figure~\ref{fig:D5_D10}), it is worth remarking that NS kinetic heating limits from muon scattering provide two extra orders of magnitude reach on the cutoff scale, $\Lambda$, than operators L1 -- L4, reaching $\Lambda \sim 10^6 \MeV$.  This occurs because L5 -- L10  are not suppressed by the lepton Yukawas. 
In addition, bounds from DM-neutron and DM-proton elastic scattering stem from one-loop induced couplings to quarks via photon and Z boson exchange between leptons and quarks. Consequently, their cross sections are less suppressed than those of L1 -- L4. 
Once again, we find that DM-muon scattering in NSs provide the most constraining upper limit. However, for some interactions (particularly L5 and L6) bounds from DM-nucleon scattering or SI nuclear recoil direct detection experiments become more competitive in the large mass region than for the scalar operators.  

As mentioned in Section~\ref{sec:loops}, operators L7 and L8, couple to quarks at one loop level only through the SM Z boson exchange, and the strength of these interactions is suppressed by either $q
_{tr}^2/M_Z^2$ or $m_\ell^2/M_Z^2$. Nevertheless, upper limits from DM-nucleon elastic scattering are approximately of the same order of magnitude as those obtained from DM-electron scattering at tree level, in particular at larger masses. In fact, they slightly surpass electron limits for DM masses $m_\chi \gtrsim 10\GeV$. 
This is due to the fact that nucleons have threshold cross sections that are more sensitive than electrons,  $\sim 4$ orders of magnitude more stringent for DM-neutron scattering. 
For operators L5 and L6, the one loop couplings to quarks are dominated by the photon contribution and, as a consequence, are less suppressed, approaching the muon bounds at larger DM masses as observed in  figure~\ref{fig:D5_D10}. 

Note  that vector-vector and axial-vector interactions, L5 and L6, showcase a particular feature, the upper limits from protons in NSs are  stronger than that of the neutrons by one order of magnitude. This occurs because the DM-proton effective coupling stems from   one-loop induced quark couplings via photon exchange.  Neutron bounds in L5 and L6, on the other hand, come solely from the Z boson contribution to the quark coupling.

Regarding DD limits for operators L5--L10, bounds from NSs are once again more stringent than Earth-based electron recoil experiments. In particular for L6, L7 and L10, operators that are velocity and momentum suppressed, DM-electron limits are more than one order of magnitude stronger than that of the future DAMIC-M experiment.
Bounds fron NSs also outperform SD limits from present and upcoming DD experiments. 
SD neutron bounds for L7 and L8 are the least stringent, because the DM-neutron effective coupling for these operators is induced at one loop by Z exchange.
In fact, SD  bounds for L7 are so weak that are not shown in figure~\ref{fig:D5_D10}, since its non-relativistic cross section, unlike that of L8,  is velocity and momentum suppressed. 
Only upper limits from SI interactions, operators L5 and L10, are competitive enough with NSs bounds in the large DM mass regime.

\section{Conclusions}
\label{sec:conclusions}

Due to their high gravitational potential, which accelerates infalling dark matter to relativistic speeds, neutron star kinetic heating is expected to provide stringent constraints on dark matter couplings to quarks and leptons, regardless of the type of the interaction. The DM capture process can heat NSs to temperatures within the reach of future infra-red telescopes, provided that the compact star is sufficiently old, faint and isolated. We have examined this kinetic heating in the context of leptophilic DM models, where fermionic DM couples primarily to leptons and DM couplings to quarks arise only at loop level. 

Neutron stars are mainly made of strongly degenerate neutrons. Nevertheless, inverse beta decay equilibrium allows protons, electrons and muons to also be present, though in lower fractions. Their precise abundances vary with the NS density profile and, together with other microscopic properties of NSs which exhibit radial dependence, cannot be determined without assuming a specific NS equation of state (EoS). Therefore, we have analysed the effect of varying the EoS on the cross section that maximises the kinetic heating, the \emph{threshold cross section}, for each particle species and chosen the most conservative NS benchmark model. 

We have used an Effective Field Theory (EFT) approach to describe DM-lepton scattering interactions and to calculate the loop induced couplings to quarks. We were thus able to derive kinetic heating upper bounds due to maximal DM capture in NSs, assuming that a single collision with a particle in a NS ($n$, $p$, $e$ or $\mu$) is required for a DM particle to become gravitationally bound to the star. 
We have found that the observation of a NS with temperature $\sim$ 1700 K can place strong upper limits on DM that couples only to leptons at tree level. The best sensitivity is obtained with DM-{\it muon} scattering, which has greater sensitivity than all other techniques across almost the entire mass range, for all types of interactions.

For capture by DM-{\it electron} scattering, neutron stars would provide significantly more constraining power than current Earth-based electron-recoil direct detection experiments, which provide only modest upper limits in the sub-GeV mass range; For some operators, future silicon based electron recoil experiments with large exposure can approach the NS electron scattering sensitivity in that same sub-GeV mass range.
For large DM mass, $m_\chi \gtrsim 10\GeV$, xenon based experiments searching for spin independent nuclear-recoil interactions can provide the strongest limits on vector-vector and tensor-tensor operators due to loop-induced interactions with nucleons. 
For the remaining operators, NS kinetic heating bounds from DM capture via scattering from either neutrons and protons, though arising at loop level, tend to be more stringent than limits from electrons. The exact range at which this occurs depends on the interaction and on the mediator of the loop-induced coupling. The latter bounds are surpassed only by the NS sensitivity to DM-{\it muon} scattering. This is especially so for the scalar and pseudo-scalar operators whose EFT couplings depend on the lepton Yukawa. The power of the muon technique is reduced at $m_\chi \sim 100\MeV$ due to Pauli blocking, yet is still more constraining than alternative approaches. 

In conclusion, muons, despite being the least abundant species in a conventional neutron star, can set the strongest bounds on leptophilic DM regardless of the type of the interaction, over several orders of magnitude of dark matter mass. We have considered the most conservative bounds on the capture of leptophilic dark matter in neutron stars, for which the stars are heated to temperatures of 1700 K, resulting in blackbody radiation in the near infra-red, within the reach of the James Webb Space Telescope (JWST). Finally, it is worth noting that colder and heavier NSs can provide even more stringent kinetic heating limits for scattering on any of the available particle species in a NS and, in the muon scattering case, can outperform terrestrial direct detection bounds over the whole dark matter mass range analysed here.

\section*{Acknowledgements}
NFB and SR were supported by the Australian Research Council.

\appendix

\section{Suppression factors on the threshold cross section}
\label{sec:suppfact}

Neglecting geometrical factors, the threshold cross section can be defined as, 
\begin{eqnarray}
    \sigma_{th} \langle n_i \rangle R_{\star} &\sim& 1\label{eq:thdef},\\
    \sigma_{th} &\sim& \frac{1}{\langle n_i \rangle R_\star} \sim \frac{R_\star^2}{N_i},
\end{eqnarray}
where $R_\star$ is the NS radius, and $n_i$ the NS baryon (lepton) number density. 

However, when the particle $i$ has a non-zero Fermi momentum, the threshold cross section (i.e., the value for which the cross section becomes large enough to achieve the geometric limit) changes. To determine the threshold cross section in this case, we can rewrite  the previous equation as
\begin{equation}
    \int_{-1}^{1} d\cos\theta \frac{d\sigma}{d\cos\theta} R_\star \langle n_i \rangle  \sim 1, 
\end{equation}
and now include the fact that not all baryons (leptons) are available for scattering as 
\begin{eqnarray}
    \int_{-1}^{1} d\cos\theta \frac{d\sigma}{d\cos\theta} R_\star \left\langle n_i \int_{(1-k(\cos\theta))p_F}^\infty dp \, n(p) \right\rangle &\sim& 1.
        \label{eq:intnp}
\end{eqnarray}
In the limit $T\rightarrow0$, the number density of states for a Fermi-Dirac distribution is given by 
\begin{eqnarray}
    n(E)dE &=& \frac{3}{2\mu_i^{3/2}}\sqrt{E} \, \Theta\left(\mu_i-E\right)dE, 
\end{eqnarray}
where $\mu_i$ is the chemical potential of the species $i$. Rewriting this in terms of the Fermi momentum and the momentum of the particle, in the classical case, we obtain
\begin{eqnarray}
    n(p)dp &=& \frac{3p^2}{p_{F,i}^3}\Theta\left(p_{F,i}^2-p^2\right)dp.
\end{eqnarray}
Now, we calculate the fraction, $f$, of particles that have a momentum larger than $p_{F,i}-q_{tr}=p_{F,i}\left(1-k\right)$:
\begin{eqnarray}
    f &=& \int_{(1-k)p_{F,i}}^\infty n(p)dp = k(3-3k+k^2)\label{eq:suppr}.
\end{eqnarray}
Then expression \ref{eq:intnp} can then be written as 
\begin{eqnarray}
     \int_{-1}^{1} d\cos\theta \frac{d\sigma}{d\cos\theta} R_\star \left\langle n_i  k(3-3k+k^2) \right\rangle &\sim& 1, 
\end{eqnarray}
which is always (for both constant or non-constant cross sections) well approximated by
\begin{eqnarray}
     \sigma_{th} R_{\star} \left\langle n_i \frac{\sqrt{\langle q_{tr}^2\rangle_\theta}}{p_{F,i}}\right\rangle &\sim& 1\label{eq:thmu},
\end{eqnarray}
where $q_{tr}$ is the momentum transfer and $\langle \rangle_\theta$ indicates average over angles. 
Therefore, the effect of the Fermi momentum suppression is to move the threshold cross section to larger values, rather than changing the value of the geometric capture rate. This result matches that of ref.~\citep{Baryakhtar:2017dbj}. 

As most of the mass is concentrated in the NS core (see Table~\ref{tab:eos}), and the Fermi momentum is approximately flat in the core (see right panel of Figure~\ref{fig:YipF}), we would expect this condition to be 
\begin{eqnarray}
     \sigma_{th} \frac{N_i}{\pi R_{core}^2} \frac{\sqrt{\langle q_{tr}^2\rangle_\theta}_{core}}{p_{F,i}^{core}} &\sim& 1.
      \label{eq:thcorePauliblock}     
\end{eqnarray}

\section{Elastic scattering cross sections}
\label{sec:xsec}
In Table~\ref{tab:operators}, we have listed the DM-lepton differential cross sections at high energy, expressed in terms of the Mandelstam variables $s$ and $t$. To compute the elastic scattering cross sections for relativistic DM, we will need the squared centre of mass energy 
\begin{equation}
s = \frac{m_\chi^2}{\mu^2}\left(1+\mu^2+\frac{2\mu}{\sqrt{B}}\right), 
\end{equation}
the minimum and maximum momentum transfers
\begin{eqnarray}
t_{max} &=& 4m_{\chi }^2 \frac{1-B}{B\left(1+\mu^2+\frac{2\mu}{\sqrt{B}}\right)},\\
t_{min} &=& 0,
\end{eqnarray}
and the Jacobian
\begin{eqnarray}
\frac{d\cos\theta}{dt} &=& B\frac{1+\mu ^2+\frac{2\mu}{\sqrt{B}}}{2\left(1-B\right) m_{\chi }^2},
\end{eqnarray}
where
\begin{equation}
\mu = \frac{m_\chi}{m_\ell}.    
\end{equation}

In Section~\ref{sec:loops}, we have derived the loop level DM-quark couplings induced by the tree-level DM-lepton couplings. In order to calculate the DM scattering cross section off neutrons, $n$, and protons, $p$, we need to compute the couplings at the nucleon level, taking into account the nuclear form factors. The squared couplings read, 
\begin{eqnarray}
C_N^S &=& \left[ \sum_{q=u,d,s}G_q^{1,2} \frac{m_N}{m_q}f_{T_q}^{(N)}+\frac{2}{27}f_{T_G}^{(N)}\left( \sum_{q=c,b,t}G_q^{1,2} \frac{m_N}{m_q} \right)\right]^2,\\
C_N^P &=& \left[ \sum_{q=u,d,s} \frac{m_N}{m_q}\left(G_q^{3,4}-\sum_q  G_q^{3,4}\frac{\overline{m}}{m_q}\right)\Delta_q^{(N)}\right]^2,\\
C_p^V &=& \left[2 G_u^{5,6} + G_d^{5,6}\right]^2,\qquad C_n^V = \left[G_u^{5,6} + 2 G_d^{5,6}\right]^2, \label{eq:veccoop}\\
C_N^A &=& \left[ \sum_{q=u,d,s} G_q^{7,8} \Delta_q^{(N)}\right]^2, \label{eq:axialcoup} \\
C_N^T &=& \left[ \sum_{q=u,d,s} G_q^{9,10} \delta_q^{(N)}\right]^2,
\end{eqnarray}
where $N$ stands for nucleon, $n$ or $p$, $G_q$ is the coupling to a given quark, $\overline{m}\equiv(1/m_u+1/m_d+1/m_s)^{-1}$ and $f_{T_q}^{(N)}$, $f_{T_G}^{(N)}$, $\Delta_q^{(N)}$ and $\delta_q^{(N)}$ are the hadronic matrix elements, determined either experimentally or by lattice QCD simulations. 
Finally the differential cross sections are obtained by substituting these coefficients in the expressions given in Table~1 of ref.~\cite{Bell:2018pkk}, with $\mu=\frac{m_\chi}{m_N}$. 


\label{Bibliography}

\lhead{\emph{Bibliography}} 

\bibliography{Bibliography} 

\end{document}